\documentclass{agujournal2019}
\usepackage{url} 
\usepackage{lineno}
\usepackage[inline]{trackchanges} 
\usepackage{soul}
\usepackage{comment}
\usepackage{float}
\usepackage{array}

\draftfalse

\journalname{ }

\begin{document}

\title{First Observation of Multiple Very-Near-Earth Reconnection Events During a Single Storm Main Phase}

\authors{Fekireselassie Beyene \affil{1}, Vassilis Angelopoulos, \affil{2},Christine Gabrielse \affil{1}, Miyoshi Yoshizumi \affil{3}, Iku Shinohara \affil{4}, 
Shoichiro Yokota \affil{5}, Satoshi Kasahara \affil{6}, Kunihiro Keika \affil{6}, 
Tomoaki Hori \affil{3}, Yasumasa Kasaba\affil{7}, Yoshiya Kasahara \affil{8}, 
Ayako Matsuoka \affil{9}, Mariko Teramoto\affil{10}, Kazuhiro Yamamoto \affil{3}}

\affiliation{1}{The Aerospace Corporation}
\affiliation{2}{University of California, Los Angeles}
\affiliation{3}{Nagoya University}
\affiliation{4}{Japan Aerospace Exploration Agency}
\affiliation{5}{Osaka University}
\affiliation{6}{University of Tokyo}
\affiliation{7}{Tohoku University}
\affiliation{8}{Kanazawa University}
\affiliation{9}{Kyoto University}
\affiliation{10}{Kyushu Institute for Technology}

\correspondingauthor{Fekireselassie Beyene}{fik.beyene@aero.org}
\correspondingauthor{Vassilis Angelopoulos}{Vassilis@ucla.edu}



\begin{keypoints}
\item Multiple VNERX events observed during single storm main phase for the first time 
\item Injections at and earthward of geosynchronous observed along with each VNERX event for the first time, suggesting VNERX is geoeffective
\item Observations during a fortuitously long residence near pre-midnight neutral sheet reveals importance of location in storm driver detection

\end{keypoints}

\begin{abstract}
For the first time, this paper presents three very-near-Earth reconnection (VNERX) events observed within the same 12-hour-long storm main phase. The THEMIS inner probes observed the hallmarks of three episodes of tailward retreating x-lines positioned between magnetic local time (MLT) 23-24 and radial distance 12-13 Earth radii $(R_E)$. The events occurred within a thin current sheet, $<1$ $R_E$ thick. Simultaneously, dispersionless energetic particle injections above 10s of keV and magnetic field dipolarizations were observed near and earthward of geosynchronous altitude by the KOMPSAT and Arase satellites. Arase observed earthward flow bursts at or below geosynchronous altitude via $\mathbf{E}\times\mathbf{B}$ enhancements, suggesting VNERX ejecta proceed below geosynchronous orbit. These observations demonstrate that VNERX events, which predominantly occur during the storm main phase, can be frequent and essential for driving injections that can effectively power the ring current. However, they can be observed only at the pre-midnight sector, close to the neutral sheet.

\end{abstract}

\section{Introduction}

Geomagnetic storms are characterized by enhanced magnetospheric convection and energy transfer, resulting from strong solar wind-magnetosphere interactions \cite{Gonzalez_94}. Dayside reconnection at the low-latitude magnetopause facilitates the deposition of solar wind energy into the magnetosphere. This energy is then redistributed across several current systems, altering the topology and intensity of the global magnetic field. During such storms, up to 70\% of the incoming solar wind energy channels into the inner-magnetospheric toroidal current, known as the ring current \cite{Li_12}. The remaining energy primarily dissipates through ionospheric joule heating, auroral precipitation, or is ejected through the dayside by magnetopause shadowing or tailward by plasmoids \cite{Poudel_19,Lu_98,Gonzalez_89}. The ring current is comprised predominantly of westward-propagating ions, typically in the few keV to tens of keV energy range \cite{Daglis_99}, which are thought to originate from the plasma sheet \cite{Fok_96, Keika_13}. However, the precise mechanisms responsible for the delivery and energization of these particles remain uncertain.

During geomagnetic storms, earthward convection in the night-side magnetosphere, known as the magnetotail, is enhanced \cite{Wolf_97}. Previous studies hypothesized that the duskward solar wind electric field penetrates the magnetotail, uniformly increasing the cross-tail electric field and driving earthward plasma transport through enhanced $\mathbf{E} \times \mathbf{B}$ drift in the plasma sheet \cite{Burton_75}. However, later studies demonstrated that the magnetosphere could not sustain such a steady-state convection pattern, as it would lead to an over-inflated inner magnetosphere and a stagnation point in the near-Earth tail region \cite{Erickson_Wolf_80, Hau_80, Lemon_04}. Using the Rice Convection Model-Equilibrium (RCM-E), a plasma convection model of the inner magnetosphere with a self-consistent magnetic field governed by $\mathbf{J} \times \mathbf{B} = \nabla P$ force balance, \citeA{Lemon_04} showed that the interchange instability enabled distant magnetotail flux tubes with reduced entropy content to move into the ring current. These low-entropy flux tubes, called plasma bubbles \cite{Pontius_Wolf_90}, exhibit stronger magnetic fields, lower plasma densities, and faster flow speeds than the surrounding plasma \cite{Chen_Wolf_93, Birn_09}. Convection models and coupled global MHD simulations suggest that these plasma bubbles are the primary drivers of plasma injection into the ring current \cite{Yang_15, Pembroke_12}.

Observations suggest that earthward magnetic flux transport in the magnetotail is primarily composed of localized flow enhancements, known as bursty bulk flows \cite<BBFs,>{Angelopoulos_92}. BBFs are signatures of plasma bubbles convecting earthward. However, observationally, BBFs cannot explain the majority of ring current enhancement, as demonstrated by the lack of strong correlation between these meso-scale flows and geo-synchronous injections \cite{Ohtani_06,Takada_06,Runov_21}. BBFs were previously thought to predominantly originate in the near-Earth tail ($>$ 20 $R_E$), presumably from reconnection sites that form there \cite{McPherron_11,Angelopoulos_94}. This location poses a concerning issue about the geoeffectiveness of BBFs, given that the plasma bubble entropy from bubbles that initiate at ~20 $R_E$ are not always low enough to allow the bubble entry to the ring current location. A viable alternative for ring current development is reconnection deep within the inner-magnetosphere. These reconnection sites would create plasma bubbles in the inner-magnetosphere with lower entropy than those formed near 20 $R_E$, making their propagation and penetration to the ring current feasible.

Recent observations have identified storm-time very-near-Earth reconnection (VNERX) events in the magnetotail, which are proposed as a potential mechanism for storm-time ring current development \cite{Angelopoulos_20, Beyene_Angelopoulos_24}. These events occurred within R = 14 $R_E$ and exhibited strong electric fields, suggesting reconnection here can efficiently accelerate low-energy particles to ring current energies. \citeA{Beyene_Angelopoulos_24} estimated an average VNERX occurrence rate between 1 and 2 VNERX events per 16 hours, the typical storm main phase duration, concluding that these could significantly contribute to the storm-time main phase ring current.

\citeA{Beyene_Angelopoulos_24} suggested that VNERX sites occur within a thin ($<$ 1 $R_E$) current sheet. Because the observing satellite must be within the narrow current sheet to make the required measurements, VNERX events are elusive and challenging to detect, complicating efforts to estimate their occurrence rate. In this study, we use the Time History of Events and Macroscale Interactions during Substorms \cite<THEMIS,>{Angelopoulos08_B} inner probes to present the first observation of multiple VNERX occurring during one storm's main phase, demonstrating for the first time what could only be inferred by \citeA{Beyene_Angelopoulos_24} regarding occurrence rates. We also demonstrate for the first time large particle injections observed near and even earthward of geosynchronous orbit around the same time, providing strong evidence of the VNERX geoffective ability.  During a 10-hour period in the main phase of the November 05, 2023, storm, three VNERX events were observed by at least one of the three THEMIS spacecraft (THA, THD, and THE). The Geostationary Korea Multi-Purpose Satellite – 2A \cite<GEO-KOMPSAT-2A, >{Kim_21} and Japanese Arase \cite{Miyoshi_18} satellites observed sudden increases in ion and electron energy fluxes that spatially and temporally coincided with the observed VNERX events, demonstrating their impacts at and below geosynchronous altitudes. 

The prior work suffered from low count statistics and was therefore unclear whether the actual VNERX occurrence rate was lower than estimated but amplified due to the poor statistics or truly representative of reality. The main issue was that the work could not establish if VNERX events occurred continuously in the pre-midnight sector and their observed occurrence rate is low due to the scarcity of satellite residence time in that location. In this work, the fortuitous timing and location of THEMIS resulted in the observation of multiple VNERX events in a single storm, demonstrating that the VNERX events are repeatable. Furthermore, this study find the VNERX occurrence rate is consistent with the prior estimate illustrating that VNERX events are not rare for storms.

\section{Data Used}
This study uses one-minute solar wind 3-D magnetic field data from the ACE and WIND magnetometers \cite{Smith98, Lepping95}. The one-minute solar wind density and velocity data come from the Solar Wind Electron, Proton, Alpha Monitor \cite<SWEPAM,>{McComas98} and WIND Solar Wind Experiment \cite<SWE,>{Ogilvie95}. All solar wind data have been time-shifted to the nose of Earth's bow shock. The three-second magnetic and electric field data of the inner-magnetosphere come from the THEMIS Fluxgate Magnetometer \cite<FGM,>{Auster08} and Electric Field Instrument \cite<EFI,>{Bonnell08} respectively. The ion and electron energy fluxes come from the Electrostatic Analyzer \cite<ESA,>{McFadden08} and Solid-State Telescope \cite<SST,>{Angelopoulos08} covering the energy ranges from 10 eV to 25 keV and 30 keV to 6 MeV, respectively. To monitor the geosynchronous particle fluxes and magnetic field evolution, we use one-minute particle data in the 100 keV to 3 MeV range from the KOMPSAT Particle Detector \cite<PD,>{Seon20} and one-second magnetic field data from the Service Oriented Spacecraft Magnetometer \cite<SOSMAG,>{Magnes20}, both datasets interpolated to match the THEMIS time cadence. To monitor ion fluxes at altitudes lower than geosynchronous, we use ion flux data from the Medium-Energy Particle Experiment \cite<MEP-i, >{Yokota_17} onboard Arase, covering 10 keV/q - 180 keV/q, interpolated to be in sync with the THEMIS data. We also analyze the magnetic and electric fields at these distances using the Magnetic Field Instrument \cite<MGF,>{Matsuoka_18} and Electric Field Detector \cite<EFD,>{Kasaba_17} of the Plasma Wave Experiment \cite<PWE,>{Kasahara_18} onboard Arase, resampled to match the time cadence of THEMIS. 

To determine the strength of ionospheric and ring current activity, we use magnetic indices derived from arrays of ground magnetometers. In this paper, we use the one-minute resolution SuperMAG electrojet (SME) index and the one-minute SuperMAG ring current (SMR) index  \cite{Newell12,Gjerloev_12}. Because the SMR index uses $\sim100$ stations as the basis for the compilation, the SMR index can be split into four indices, each representing the average change of the ground horizontal field in a longitudinal sector quadrant. The SMR index thus gives an enhanced spatial resolution of the ring current evolution. To supplement the SMR index, we also provide the one-minute Sym-H \cite{Iyemori_90} index, which is analogous to the SMR but uses six mid-latitude stations. Unless otherwise specified, all data are expressed in the Geocentric Solar Magnetospheric (GSM) coordinate system. 

The THEMIS electric field data utilized Despun-L-vector (DSL) coordinates. This coordinate system is defined as local to the satellite and defines a spin-plane and spin-axis.  Due to its low-inclination orbit ($<$ 10\textdegree \space from the ecliptic plane), the THEMIS DSL coordinate system aligns closely with the GSM coordinate system. On Arase only the two spin-plane components of the electric field are measured; thus, the normal-to-spin plane component is derived assuming $\bf{E} \cdot \bf{B} = 0$.

\section{Methodology: Identification of VNERX events}
\begin{figure}[H]
\noindent\includegraphics[width=1.\textwidth]{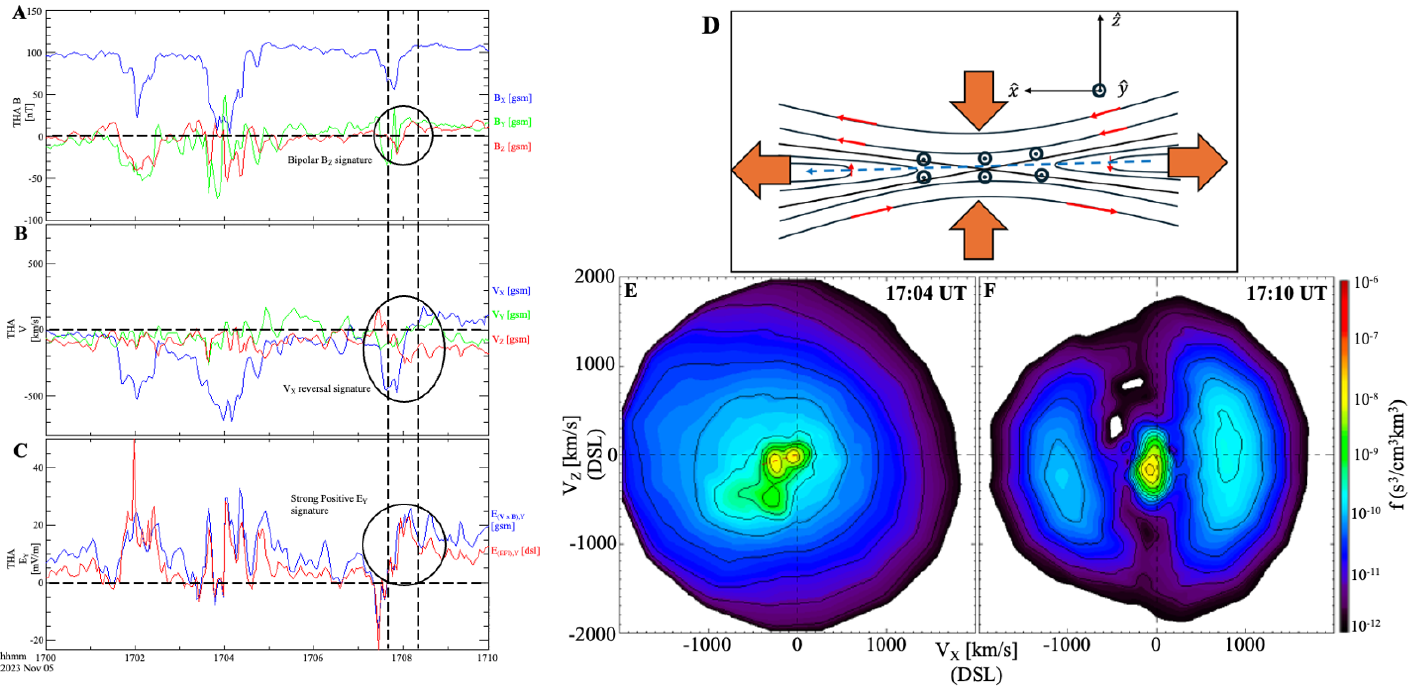}
\caption{Schematic detailing how VNERX events are inferred from THEMIS observations. A) THEMIS-A (THA) observations of the magnetic field in GSM coordinates including the bipolar BZ signature due to the tailward-traveling VNERX passage. B) THA observations of bulk ion velocity displaying the tailward (-XGSM) reconnection outflow in blue. Several instances of tailward outflows are seen at 17:02 UT, 17:04 UT, and 17:08 UT. C) The EFI-observed and estimated y-component of the electric field in red and blue respectively. D) Idealized X-line geometry of a magnetotail reconnection site. The dashed blue line denotes the trajectory of a spacecraft in the frame of a tailward traveling reconnection site. E) 2D XZ ion velocity cut in DSL coordinates from burst mode ESA data during the fast flow seen at 17:04 UT. F) 2D XZ ion velocity cut in DSL coordinates from burst mode ESA data during the earthward flows seen at 17:10 UT.}
\label{fig1}
\end{figure}

We examined the THEMIS data during the storm period November 04 - 11, 2023. At this time the THEMIS spacecraft had a highly eccentric orbit whose apogee of 14 $R_E$ was in the pre-midnight magnetotail. We define the very-near-Earth region as the inner magnetospheric region within 14 $R_E$ and reconnection events found in this region as very-near-Earth reconnection events. To find VNERX events, we employ the criteria developed in \citeA{Beyene_Angelopoulos_24}. These criteria are based on an idealized picture of an X-line reconnection site moving tailward. Figure \ref{fig1} gives a schematic that relates the data observations with our interpretation of an X-line that traverses a spacecraft. An X-line with no guide-field consists of magnetic field lines with a negative (positive) $B_Z$ component tailward (earthward) of the X-line. In Figure \ref{fig1}D, the schematic shows magnetic field lines as solid black curves with red arrows overlaid to denote the local direction of the magnetic field. The large orange arrows denote the direction of bulk plasma flow, indicating the inflow and outflow regions. Magnetic flux carried away from the X-line results in a reconnection electric field in the out-of-plane direction; in Figure \ref{fig1}D, the circled dots denote this $E_Y$ field.

Consider a VNERX site created earthward of a spacecraft. As the VNERX site travels tailward (as reconnection is shown to do, as pressure from the dipolarized field earthward of it pushes it tailward \cite<e.g.,>{Hones_85}), the spacecraft will observe first the tailward features of the VNERX site, and following the earthward features. In Figure \ref{fig1}D, the dashed blue line denotes the trajectory of a spacecraft in the frame of the traveling reconnection site. The criteria developed for tailward-traveling VNERX site identification are: (a) a bipolar (negative-to-positive) $B_Z$ signature, (b) a strong positive $E_Y$ throughout the ion diffusion region (IDR), (c) a fast tailward flow concurrent with the negative $B_Z$ interval, with earthward flows not required but, when present, positively correlated with the positive $B_Z$ interval.

We visually inspected the storm period for negative-to-positive $B_Z$ signatures using these criteria. In the inner magnetosphere, the magnetic field strength is 10s of nT and the ion density is around 0.1 to 0.5 cm$^{-3}$, giving an ion Alfvén velocity $(V_{Ai})$ around 2000 km/s. Past reports identify magnetotail reconnection sites using an ion flow threshold of $|V_X| \ge 100$ km/s \cite{Eastwood_10, Artemyev_17}. Guided by these past works, we define fast flows as those greater than $0.1*V_{Ai} = 200$ km/s and thus we kept times when the ESA-derived velocity showed a large tailward flow of $<-200$ km/s simultaneously with a negative $B_Z$. We did not require a positive flow to be observed, as reconnection sites are thin, making it unlikely to see both sides of the reconnection sites. We also require a large positive $E_Y$ within a 30-second window centered around the $B_Z$ zero-crossing time. \citeA{Rogers_19} used a threshold of $|E| >$ 10 mV/m to detect the ion diffusion region (IDR) associated with reconnection sites. Strong electric field signatures typically exist close to the neutral sheet and x-line center \cite{Eastwood_2010}, thus taking into account THEMIS may only skim the reconnection site from above or below due to the sites being embedded in an extremely narrow current sheet, we relax the \citeA{Rogers_19} criteria and set a threshold of $E_Y >$5 mV/m.  The left panels of Figure \ref{fig1} show an example of the VNERX signatures we searched for. Using our criteria, we found three VNERX events. THA observed the associated $B_Z$ zero crossings at 17:08 UT and by THD at 14:26 UT and 19:27 UT. See supplemental material Figure S2 for a summary of the three events.

As an additional check, once the events were selected, we created 2D velocity cuts of the low-energy ion phase space distribution at the instances of fast flows to verify that the flows were caused by reconnection. We look for hallmarks of reconnection in the ion dynamics: simultaneous observations of cold ion inflow and hot ion outflows. Figure \ref{fig1}E shows the 2D cut during the second fast tailward flow observed by THEMIS A during VNERX 2 at 17:04:10 UT, and \ref{fig1}F shows the fast earthward flow at 17:10:00 UT in the X-Z DSL plane. The x-axis is generally along the sun-Earth line, and the z-axis is perpendicular to the ecliptic. In Figure \ref{fig1}E, the distribution clearly shows a cold, low-velocity population flowing towards the neutral sheet in the -Z direction. We interpret this population as cold ions inflowing towards the reconnection site. The 2D cut also shows a background flow in green and light blue that is faster and more aligned in the -X direction. This population is interpreted as the fast, energized ion population outflowing from the reconnection site. In Figure \ref{fig1}F, there is a prominent population with large +$V_X$, representing the fast earthward reconnection outflows, and a -$V_X$ population representing reflected cold ions \cite{Onsager_91, Forbes_81} that have not had time to reach the reconnection site yet, indicating the observed fieldline was very recently reconnected. In Figure \ref{fig3_new} we show the chronology of observed 2D X-Z velocity distributions spanning 10 minutes from 17:00 UT to 17:10 UT. This period shows the distribution before VNERX 2 passed THA (17:00 UT - 17:07 UT) and after passage (17:09 UT - 17:10 UT). We Before the $B_Z$ reversal, THA observed a distribution that showed a prominent population that is mostly tailward (negative $V_X$) and towards the neutral sheet (negative $V_Z$, noting THA witnessed positive $B_X$ at these time), see 17:00 UT, 17:01 UT, and 17:05 UT. During the $B_Z$ reversal and for the two minutes afterwards the distribution shows two lobes that represent the fast earthward flow (positive $V_X$) and reflected tailward moving ions traveling on yet-to-be reconnected field lines and a slow inflow population. From these distributions we visually deduce that the inflowing plasma represent a colder lobe population and the outflowing population represent the hotter plasma sheet population that exists on the reconnected field lines. Supporting Information S8 and S10 present the 2D X-Z cuts for VNERX \#1 and \#3, respectively.

\begin{figure}[H]
\noindent\includegraphics[width=1.\textwidth]{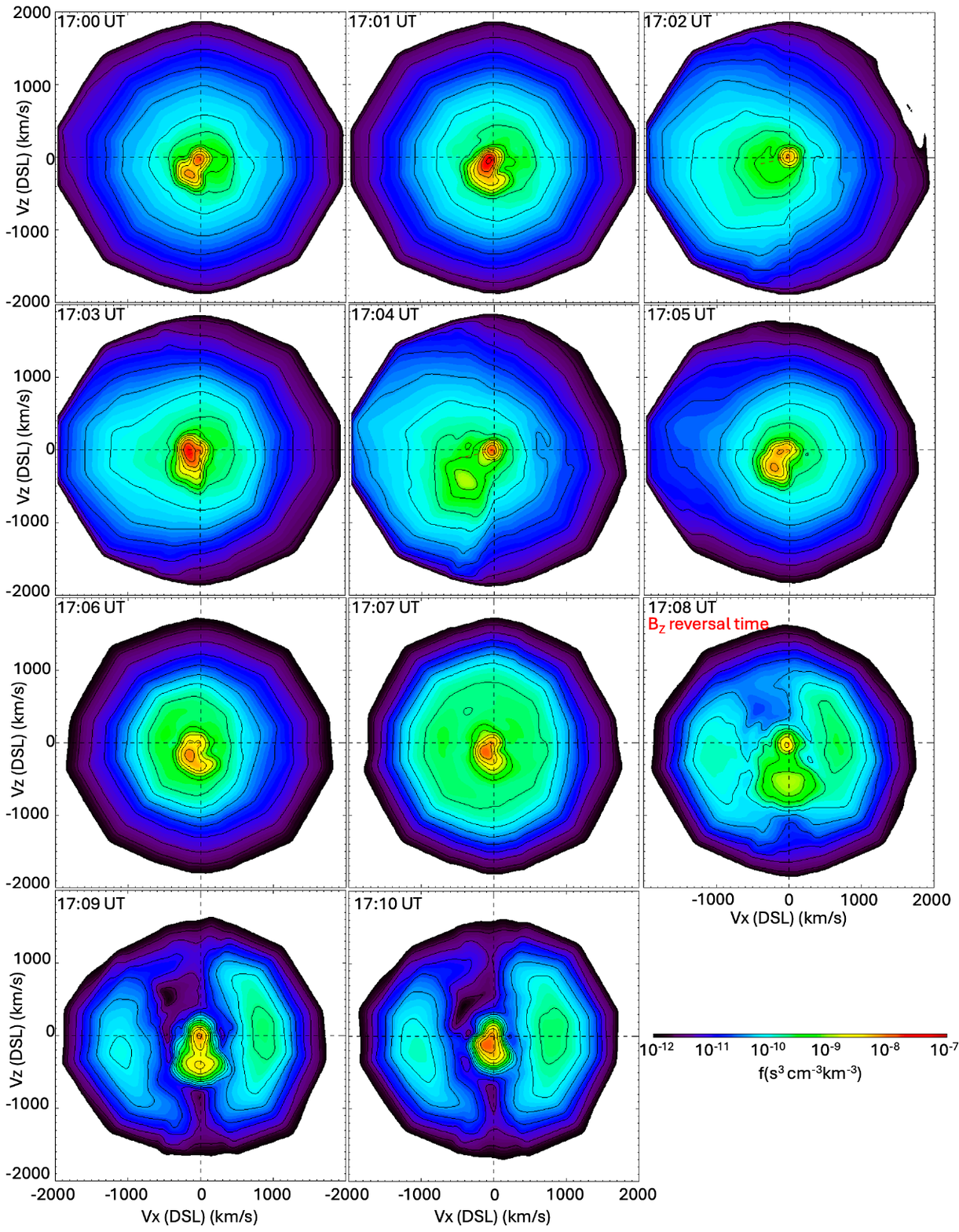}
\end{figure}

\begin{figure}[H]
\caption{2D ion velocity cuts in $V_{X,DSL}$-$V_{Z,DSL}$ space from the THA reduced mode ESA distribution. The cuts are ordered in time from 17:00 UT to 17:10 UT.}
\label{fig3_new}
\end{figure}

When identifying VNERX events, we account for neutral sheet tilt effects, which can distort the magnetic field normal component and contaminate the bipolar $B_Z$ signature by introducing a finite $B_X$ component. Using the Tsyganenko et al. (2015) TAG14 neutral sheet model, we determine the neutral sheet position in the Y = $⟨Y_{GSM}⟩$ plane for each event based on the average $Y_{GSM}$ location of THA during a one-hour window centered on the $B_Z$ zero crossing time, then rotate the magnetic field and velocity data into a local neutral sheet coordinate system ($X_{NS}$ tangent to the neutral sheet toward Earth, $Y_{NS}$ along $Y_{GSM}$, $Z_{NS}$ normal to the neutral sheet). All VNERX events maintained their bipolar $B_{Z,NS}$ signature after this rotation, confirming that neutral sheet tilt did not cause misidentification of these events.

\section{Results}
A moderately strong geomagnetic storm with a Sym-H minimum of -163 nT occurred from November 04-11, 2023. During this time, the THEMIS apogee was positioned in the pre-midnight tail. On November 5th, from 12:00 UT to 22:00 UT, during the late-stage storm main phase, THEMIS detected three distinct VNERX events. At this time, KOMPSAT observed the pre- and post-midnight inner-magnetosphere at geosynchronous orbit and recorded three dispersionless injections in the proton and electron energy fluxes. Arase completed more than one orbit during this period and observed two ion flux enhancements in the 10-180 keV range near apogee. These VNERX events exhibited classic signatures of X-line reconnection: prolonged periods of strong reconnection ($E_Y$) electric fields, $B_Z$ and $V_X$ reversals, and quadrupolar Hall ($B_Y$) magnetic fields. Specifically, during the 1-minute periods following the VNERX crossings the $E_Y$, either estimated from $\mathbf{V} \times \mathbf{B}$ or observed via EFI, was greater than 5 mV/m for at least 30 seconds. In the Supplementary Material we show our analysis of the quadrupolar $B_Y$ signature.  Figure S6 shows a scatter plot of $B_X$ vs $B_Z$ colored by $B_Y$. We find that during the periods of strong $E_Y$ the expected quadrupolar $B_Y$ behavior, positive $B_Y$ to $B_Z$ correlation during positive $B_X$ intervals and negative $B_Y$ to $B_Z$ correlation otherwise, is seen. 

The VNERX observations indicate that reconnection events recur during the storm main phase and may play a significant role in particle injections at, and below, geosynchronous altitudes during this phase. These reconnection sites are analogous to reconnection sites seen further downtail, only stronger in their electromagnetic field properties, thus capable of higher local particle energization. Three VNERX events during a single storm main phase have never been observed due to the low likelihood of satellites being positioned within the less than 1 $R_E$ thick current sheet. The fact that THEMIS was well-situated to observe three VNERX events demonstrates that (1) spacecraft position is crucial for VNERX detection, and (2) VNERX events likely occur repeatedly during the storm main phase at that location, but are often missed because they exist in such a thin current sheet. 

\subsection{Overview of Events}

Figure \ref{fig2} shows the THEMIS, KOMPSAT, and Arase orbit tracks and corresponding magnetic field and particle data during the Nov. 5th, 2023 storm main phase. In Figure \ref{fig2}A, the cross, triangle, and pentagon symbols mark the position of the spacecraft at the time the VNERX events were observed by THEMIS. The THEMIS inner probes traveled earthward and dawnward from apogee located near 14 $R_E$ at 23 MLT, KOMPSAT traveled from 21 MLT to 6.5 MLT, and Arase completed an orbit starting near apogee at X = -6 $R_E$. Figure \ref{fig2}B, C, and D show the magnetic field recorded by THA, THD, and THE, respectively. The THEMIS spacecraft observed three periods of highly variable magnetic field: 14:00 - 15:30 UT, 17:00 - 18:00 UT, and 19:00 - 20:00 UT. During these periods THEMIS observed instances of $B_X \sim 0$, suggesting that THEMIS was located near the neutral sheet and capable of observing VNERX. Using the TAG14 neutral sheet model \cite{Tsyganenko_15}, we estimated the distance of THEMIS to the neutral sheet during 10-minute windows surrounding the VNERX events. We found that the observing THEMIS spacecraft was always within 1 $R_E$ of the neutral sheet. In Figure \ref{fig2}, the dashed red vertical lines mark when tailward-propagating VNERX passed a THEMIS satellite (14:26 UT, 17:08 UT, and 19:27 UT), and the black vertical lines mark when KOMPSAT observed dispersionless electron injections (14:00 UT, 17:20 UT, 19:19 UT). We note that VNERX \# 3 showed fast flows three to seven minutes prior to the $B_Z$ reversal indicative of the VNERX passage. However, fast tailward flows of 500 km/s apparent in this event occurred for several minutes and punctuated with instances of negative BZ suggesting that the spacecraft (THD) was witnessing the tailward jets of an Earthward reconnection site. Note we suggest that the fast tailward flows can be seen well ahead of the BZ reversal period, because the tailward flows travels at 100s of km/s while the reconnection site itself, driven by tailward pressure gradients, may only travel at 100 km/s \cite{Russell_McPherron_73}. At the time of the quoted VNERX crossing (19:27 UT), THD was outside the plasma sheet, and thus incapable of witnessing the fast tailward flows that likely occurred at the neutral sheet. 

In Table \ref{Table1} we provide the times of the VNERX observations, the KOMPSAT dispersionless injections, and the Arase ion flux enhancements. The vertical red-dashed lines in Figure \ref{fig2} correspond to the passage of the VNERX point/line by the satellite during its tailward progression. However, VNERX must have started earlier, closer to the Earth, which in part contributes to some injections preceding the VNERX tailward motion past THEMIS (Event \#1 and Event \#3). Likely many VNERX events occur during this several-hour interval but the exit of the satellite from the close proximity to the neutral sheet (as evidence during instances of strong $|B_X|$) prevents detection of the activity. We were lucky to observe clear signatures of reconnection when we did and were able to identify specific times that they progressed tailward. However, as evidenced by the tailward flows which last a lot longer than the events we studied, the VNERX activity was ongoing and the times when it intensified abruptly enough to register another injection at geosynchronous cannot be deduced from THEMIS tail observations. In other words, THEMIS observations of tail activity as severely aliased by plasma sheet flapping, causing exit of the spacecraft from the thing ion-scale reconnection region where the activity can be registered as tailward-earthward flows and corresponding bipolar $B_Z$.

\begin{table}[H]
    \centering
    \begin{tabular}{|>{\centering\arraybackslash}p{0.33\linewidth}|>{\centering\arraybackslash}p{0.33\linewidth}|>{\centering\arraybackslash}p{0.33\linewidth}|}\hline
         aliased VNERX crossing observation time&  KOMPSAT dispersionless injection time& Arase ion flux enhancement time\\\hline
         14:26 UT&  14:00 UT& -\\\hline
         17:08 UT&  17:20 UT& 17:08 UT \\\hline
 19:27 UT& 19:19 UT&19:17 UT\\ \hline
    \end{tabular}
    \caption{Table of times of VNERX crossing observations and particle obserations at geosynchronous and lower altitudes.}
    \label{Table1}
\end{table}

Figure \ref{fig2}E presents the magnetic field measured by Arase in the same format as THEMIS. Arase observed two instances of magnetic field dipolarization at 17:08 UT and 19:17 UT, corresponding to VNERX 2 and 3. Arase was near perigee and unable to observe any related signatures to the VNERX 1 at 14:26 UT. Figure \ref{fig2}F and G display the ion and electron number flux. Two instances of particle flux enhancement across several energies are clearly seen in both panels. These instances occur shortly before or at the same time as the VNERX site crossing THEMIS. Figure \ref{fig2}H shows the magnetic field observed by KOMPSAT. $B_X$ was negative for the entire period, indicating KOMPSAT was below the neutral plane. The strong $B_X$ and small Bz suggest the plasma sheet encroached very near-Earth. KOMPSAT observed three classical dipolarizations \cite<see,>{Sergeev_98, Liu_16} at 14:14 UT, 17:20 UT, and 19:19 UT. Figure \ref{fig2}I and J show the proton and electron number flux observed by KOMPSAT, respectively.  Enhancements in these number fluxes are seen near the times of the observed KOMPSAT dipolarizations. These three dipolarization and injection events correspond to the timings of the three VNERX observed by THEMIS. Figure \ref{fig2}K shows the SMR and Sym-H indices. The red, green, and blue curves are the deviations of the horizontal magnetic field in the MLT sector wedges centered on MLT 18 hr, 06 hr, and 00 hr, respectively. The black curve gives the average deviation across all MLTs. 

\begin{figure}[H]
\noindent\includegraphics[width=1.\textwidth]{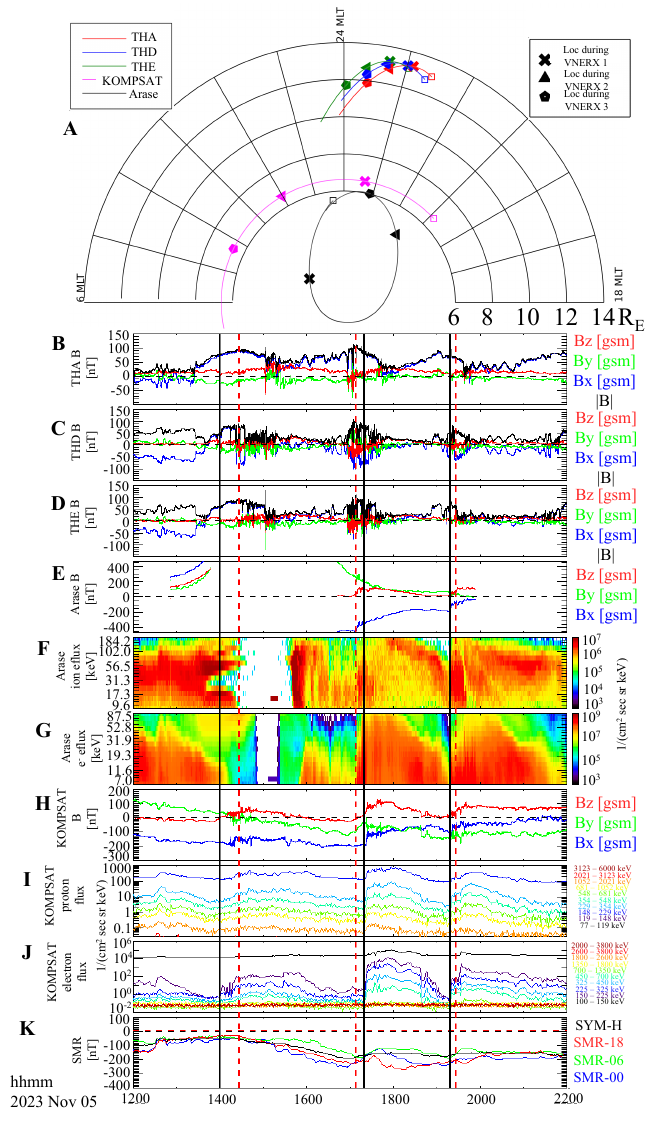}
\end{figure}

\begin{figure}[H]
\caption{X-Y GSM positions of Arase, KOMPSAT, and THEMIS inner probes in $R_E$ and corresponding magnetic field and particle observations from November 05, 2023 12:00 - 22:00 UT. A) X-Y GSM orbit tracks of Arase, KOMPSAT, and THEMIS inner probes. X, triangle, and pentagon denotes the positions of the spacecraft during the time when a $B_Z$ reversal (south-to-north) associated with a VNERX site passage was observed by one of the THEMIS spacecraft. The black, pink, red, blue, and green traces refer to Arase, KOMPSAT, THA, THD, and THE respectively. For panels B - K the three vertical black lines denote times when KOMPSAT observed dispersionless injections in the electron energy fluxes and the red vertical dashed lines denote the time a VNERX site passed THEMIS, determined via an observed $B_Z$ and $V_X$ reversal (14:26:00 UT, 17:08:00 UT, and 19:24:13 UT). B) Magnetic field components observed at THA. C) Magnetic field components observed at THD. D) Magnetic field components observed at THE. E) Magnetic field components observed at Arase. F) Ion number flux observed by Arase. G) Electron number flux observed by Arase. H) Magnetic field components observed at KOMPSAT. I) Proton number fluxes from 100 keV - 3 MeV observed at KOMPSAT. J) electron number fluxes from 100 keV - 3 MeV observed at KOMPSAT. K) SMR index. The blue, green, and red curves represents the magnetic deflection occurring in the quarter longitudinal MLT wedge centered on MLT = 0, 6, and 18 hr respectively and the black curve represents the north-south deflection averaged over all MLTs.}
\label{fig2}
\end{figure}

Supplemental Figure 1 describes the solar wind environment with an interplanetary magnetic field demonstrative of a coronal mass ejection \cite<ICME,>{Lepping_90} that intersected Earth's magnetosphere. Coinciding with the increase in solar wind dynamic pressure at 09:09 UT (SI Figure 1C), all ring current indices sharply increased above 0 nT. The increase in the indices is an example of a storm-sudden commencement \cite<SSC,>{Mayaud_75}, where a substantial and rapid increase in solar wind dynamic pressure compresses the magnetosphere, and the magnetopause currents respond by increasing to return the system to global pressure balance. The increased magnetopause currents result in an average positive deviation of the ground equatorial magnetic field. The sustained positive SMR and Sym-H excursion following the SSC is known as the storm initial phase \cite{Gonzalez_94}. The indices turned negative at 10:00 UT. This feature marks the start of the storm main phase \cite{Gonzalez_94}, where the intensification of the ring current decreases the horizontal component of the low-latitude ground magnetic field. The storm exhibited a common two-step structure \cite{Kamide_98} caused by a negative solar wind $B_Z$ both in the sheath ahead of the ICME and within the ICME itself. The SMR indices allow us to investigate the azimuthal dependence of ring current activity. In both minima at 11:10 UT and 18:10 UT SMR 18 showed the lowest value, indicating the ring current was most intense in the pre-midnight sector. 

\begin{figure}[H]
\noindent\includegraphics[width=\textwidth]{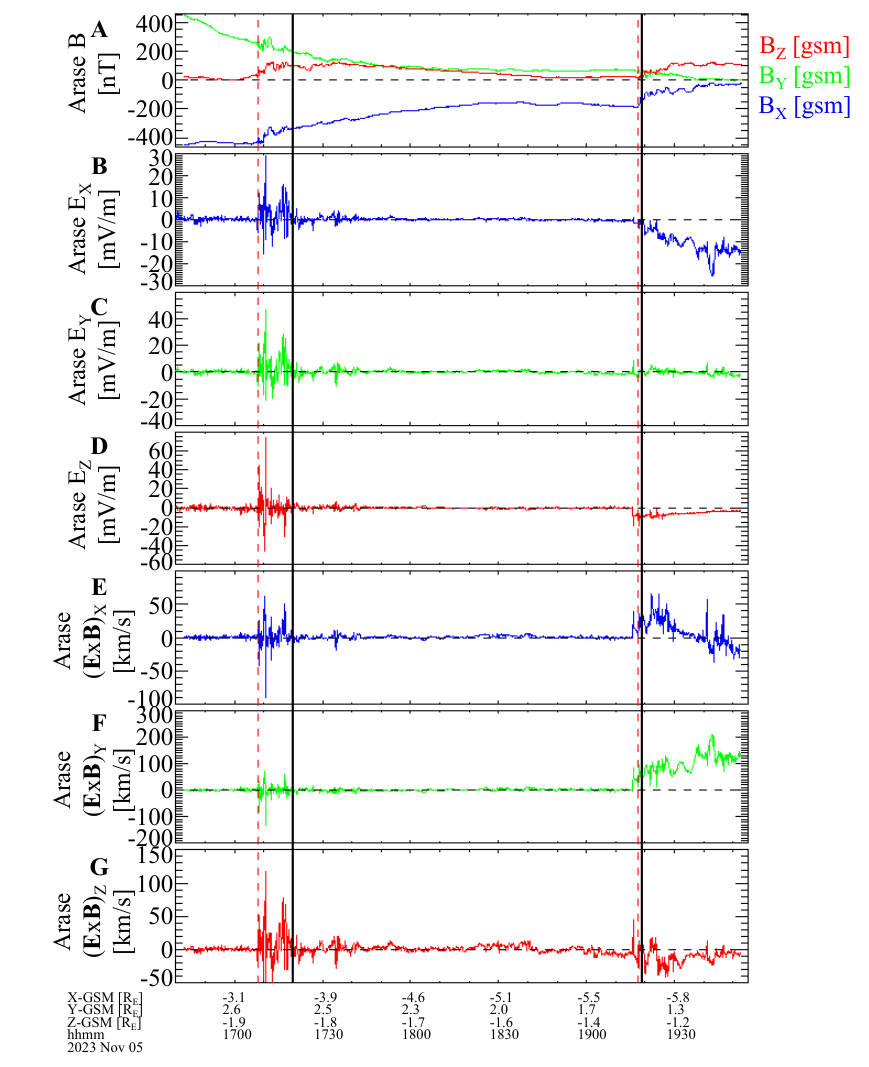}
\caption{The magnetic and electric fields measured by Arase and inferred ExB drift for VNERX \#2 and \#3. The black vertical lines denote times when KOMPSAT recorded a dispersionless injection and the red dashed vertical lines denote times when Arase saw an ion energy flux enhancement. A) Magnetic field measured by Arase. B) x-component of electric field measured by Arase. C) y-component of electric field measured by Arase. D) z-component of electric field measured by Arase. E)x-component of ExB drift. F)y-component of ExB drift. G) z-component of ExB drift.}
\label{fig3}
\end{figure}

Using the electric and magnetic field observations from Arase, we can determine how bulk plasma flows at low altitudes during these VNERX events. Figure \ref{fig3} shows the electromagnetic fields observed by Arase and the derived $\mathbf{E}\times \mathbf{B}$ drift velocity. Figure \ref{fig3}A shows the magnetic field measured by Arase as described in the previous paragraphs. Figure \ref{fig3}B, C, and D show the x-,y-, and z-component of the electric field that Arase measured by the EFD. The z-component was derived assuming $\mathbf{E} \cdot \mathbf{B} = 0$. 

The red vertical dashed lines denote times Arase observed ion number flux enhancements. These panels demonstrate that the electric field intensified with the observed enhancement of energetic particle number flux. Figure \ref{fig3}E, F, and G show the x-,y-, and z-component of the $V_d = \mathbf{E} \times \mathbf{B}$ drift. The drifts suddenly appeared at the start of the ion injection and remained strong for the subsequent 10 minutes. The z-component drift frequently reached over 50 km/s concomitant with positive x-component drift of up to 60 km/s and decreasing $B_X$ magnitude, implying earthward flows contracting toward the neutral plane. The drift and magnetic field signatures indicate that Arase observed earthward-traveling dipolarizing flux bundles \cite{Liu_16, Liu_14}. These observations imply that the earthward flow bursts driven by VNERX may proceed deep into the inner magnetosphere beyond geosynchronous. Although \citeA{Runov_22} demonstrated an injection at geosynchronous orbit related to a single VNERX event, this is the first set of observations of deep particle penetration \textit{beyond} geosynchronous corresponding to multiple VNERX events in a single storm. Deep penetration of flows may cause significant particle energization, and these energized particles can become trapped stably for prolonged periods. The evidence for the importance of VNERX in driving a geomagnetic storm is growing.

This overview showcased three periods when THEMIS observed signatures of a tailward retreating VNERX while close to the neutral sheet. During these periods, both KOMPSAT and Arase observed magnetic field dipolarizations and KOMPSAT observed dispersionless injections, suggesting the plasma sheet was first stretched at very-near-Earth distances and then dipolarized.

\section{Discussion}

The observations in this study demonstrate that during the storm main phase, VNERX can occur as often as 1 VNERX/4 storm main-phase hours.

The THEMIS probes observed the storm's main-phase tail for the entire main phase, 12.17 hours. We treat each spacecraft as an independent observer; thus, the total inner-magnetosphere storm main-phase observation time is 36.51 hours, and the VNERX events were observed at distances of R = [-12, -14] $R_E$. If we consider residence time only when THEMIS is within 0.5 $R_E$ of the neutral sheet, around the expected size of the reconnection ion diffusion region \cite<several ion inertial lengths, 1 $\lambda_i = 1000$ km, >{Nagai_21}, the occurrence rate is 204 VNERX events/ 1000 storm main-phase hours. Figure 8 from \citeA{Beyene_Angelopoulos_24} suggests an occurrence rate of 62 - 128 VNERX events/1000 storm main-phase hours if we consider residence time within 1000 km of the neutral sheet. However, this averages over all spatial bins that contained VNERX events. To directly compare with our study, it is appropriate to reassess their occurrence rates only considering bins tailward of R=11 $R_E$ since the events in this study were observed at 12 - 14 $R_E$. Additionally, because THEMIS can observe a tailward retreating VNERX that initiated earthward of THEMIS, but cannot observe a tailward retreating VNERX that initiates tailward of THEMIS, the \citeA{Beyene_Angelopoulos_24} occurrence rate determination may have been underestimated by including bins closer to Earth. By restricting the \citeA{Beyene_Angelopoulos_24} to only those bins tailward of R = 11 $R_E$, the average VNERX occurrence rate is 200 VNERX events/1000 storm main-phase hours. Our results agree with this estimate.

We use a 1D Harris current sheet fit to estimate the current sheet thickness prior to the passage of VNERX 2. Shown in Figure S5, the current sheet thickness during this period is $ \le 1 R_E$. This estimate suggests these events are easily missed due to the thin current sheet they exist within.

We suggest that during the storm main phase, the solar wind compression not only thins the current sheet, but also moves the plasma sheet earthward. Very-near-Earth thin current sheets have recently been inferred by \citeA{Shi_24} for a substorm event. In that study, the authors used a robust magnetotail magnetic field reconstruction algorithm \cite{Stephens_19}. They demonstrated that the electron and ion isotropy boundaries near MLT = 0 approached 5 - 13 $R_E$ during the growth phase. \citeA{Stephens_25}  also showed two $B_Z$=0 loops (reconnection) \textit{inside} of geosynchronous orbit, and that the loops moved earthward with increased solar wind-magnetosphere interaction (a stronger solar wind electric field), which they pointed out is consistent with the Geotail statistics \cite{Nagai_05, Nagai_23}

The events we studied suggest that VNERX occurs predominantly in the pre-midnight sector, in agreement with the conclusions of \citeA{Beyene_Angelopoulos_24}. The duskside preference of reconnection has been reported previously for near-Earth (as opposed to Very-Near Earth) tail reconnection \cite<$\sim$20 - 30 $R_E$,>{Genestreti_14,Eastwood_10,Nagai_13,Nagai_98}. For these phenomena it have been suggested that a Hall electric field $E_Z$ develops, creating a dawnward-directed $\mathbf{E} \times \mathbf{B}$ drift, evacuating the duskside current sheet of magnetic flux and plasma \cite{Lu_18, Lu_18_b}. We suggest that this process is also applicable to VNERX events, are consistent with the thin current sheet formation and the pre-midnight location of the VNERX sites.

\section{Conclusion}

This study reports, for the first time, multiple observations of three VNERX events occurring during the same storm main phase. We also report the first observations of particle injections associated with the VNERX events earthward of geosynchronous orbit. We highlight that the frequency suggests these events are important for storm growth. We showcase the solar wind conditions during the day the VNERX events were observed and concluded that a strong solar wind dynamic pressure plays a significant role in generating VNERX, likely compressing the plasma sheet and extending it earthward. We further report, for the first time, the pre-midnight preference for VNERX, similar to near-Earth (~20 $R_E$) reconnection events.

The same statistical result on the occurrence rate of \citeA{Beyene_Angelopoulos_24} is now verified in our case study. We have shown that VNERX is a highly repeatable storm-time main phase phenomenon, well-observed if a satellite resides for a prolonged period in the pre-midnight sector and is sufficiently close to the neutral sheet. 

This study stresses that the very-near-Earth magnetotail is critical in storm development. To improve constraining the VNERX occurrence rate under a variety of solar wind conditions and their underlying physics requires more spacecraft observations of this pre-midnight sector region during storms.  Complementing these observations with inner-magnetospheric convection model results will help determine the global structure of the thin plasma sheet. More observations of this region will quantify the evolution of the very-near-Earth x-lines, what geomagnetic and solar wind drivers are essential in creating them, and better establish the transport of the energetic ion and electron distributions from the VNERX sites to the ring current.

%
%

\urlstyle{same}
\section*{Open Research Section}
THEMIS data are available through \url{http://themis.ssl.berkeley.edu}. The OMNI data, including solar wind parameters and geomagnetic indices, were obtained from the GSFC/SPDF OMNIWeb interface at \url{http://omniweb.gsfc.nasa.gov}. Science data of ERG (Arase) satellite were obtained from the ERG Science Center operated by ISAS/JAXA and ISEE/Nagoya University \cite{Miyoshi_18b},  \url{https://ergsc.isee.nagoya-u.ac.jp/data\_info/erg.shtml.en}. The KOMPSAT data are available through \url{https://swe.ssa.esa.int/hapi/data?id=spase://SSA/NumericalData/GEO-KOMPSAT-2A/}. THEMIS, Arase, KOMPSAT, and OMNI data have been imported and analyzed using corresponding plug-ins to the SPEDAS analysis platform \cite< \url{http://spedas.org} Angelopoulos et al.,>{Angelopoulos_19}. The Tsyganenko et al. (2015) neutral sheet model is available through \url{https://geo.phys.spbu.ru/~tsyganenko/empirical-models/current_sheet/neutral_sheet/}. 

\acknowledgments
The THEMIS mission is supported by NASA contract NAS5-02099, NSF grant AGS-1004736, and CSA contract 9F007-046101. The present study analyzed MGF-L2 8 sec spin-averaged data v04\_04 (10.34515/DATA.ERG-06001), MEP-i-L2 omniflux data v02\_01 (10.34515/DATA.ERG-03001), Orbit L3 v02 data (10.34515/DATA.ERG-12001), and PWE-EFD L2 data v01\_02 (10.34515/DATA.ERG-07004)

%
%

\bibliography{VNERX_references}

\newpage

\end{document}


%
%


\title{Supporting Information for "First Observation of Multiple Very-Near-Earth Reconnection Events During a Single Storm Main Phase"}
%
%

%
%

\authors{Fekireselassie Beyene \affil{1}, Vassilis Angelopoulos, \affil{2},Christine Gabrielse \affil{1}, Miyoshi Yoshizumi 
\affil{3}, Iku Shinohara 
\affil{4}, Shoichiro Yokota 
\affil{5}, Satoshi Kasahara 
\affil{6}, Kunihiro Keika 
\affil{6}, Tomoaki Hori 
\affil{3}, Yasumasa Kasaba
\affil{7}, Yoshiya Kasahara 
\affil{8}, Ayako Matsuoka 
\affil{9}, Mariko Teramoto
\affil{10}, Kazuhiro Yamamoto 
\affil{3}}


\affiliation{1}{The Aerospace Corporation}
\affiliation{2}{University of California, Los Angeles}
\affiliation{3}{Nagoya University}
\affiliation{4}{Japan Aerospace Exploration Agency}
\affiliation{5}{Osaka University}
\affiliation{6}{University of Tokyo}
\affiliation{7}{Tohoku University}
\affiliation{8}{Kanazawa University}
\affiliation{9}{Kyoto University}
\affiliation{10}{Kyushu Institute for Technology}

%
%

%

\begin{article}
\raggedbottom
%
%

\clearpage
\noindent\textbf{Contents of this file}
\begin{enumerate}
\item Text S1 to S5
\item Figures S1 to S10
\end{enumerate}

\clearpage

\noindent\textbf{Introduction}

In the supporting information we provide more detail on how we identified VNERX events, taking into account neutral sheet tilt and distinguishing the observations from other magnetotail structures that would create similar signatures in the data. We next provide a case study that details what a VNERX event looks like from the combined magnetic field, electric field, and plasma measurements from THEMIS. In concert with the Arase and KOMPSAT data, the case study overview describes a tailward passage of a VNERX site passed THEMIS and the following fast earthward particle injections observed below geosynchronous altitude. We then provide a multi-case analysis that solidifies our understanding of the VNERX site electromagnetic field structure. The statistics reveal the well-known X-line picture describes well the VNERX site geometry and that these X-lines host much stronger fields relative to their near-Earth counterparts.


\clearpage

\noindent\textbf{S1: Overview of Solar Wind Dynamics}

The solar wind that drove this storm contained periods of negative $B_Z$ and strong solar wind dynamic pressure. Figure \ref{figS1} shows the solar wind properties and global magnetospheric state via the SME and SMR/SYM-H indices during November 05, 2023. The black vertical lines denote when KOMPSAT observed dispersionless injections in the electron fluxes, and the red dashed vertical lines denote times when Arase observed simultaneous magnetic field dipolarizations and proton number flux enhancements. Figure \ref{figS1}A shows that the SME shows four peaks that start at $\sim$09:00 UT, $\sim$14:00 UT, $\sim$17:30 UT, and $\sim$19:10 UT. The start of the final three peaks line up well with the times KOMPSAT observed energy flux enhancements. Since SME is strongly correlated with nighttime auroral activity \cite{Newell11}, the peaks in SME imply intensifications of nightside auroral activity occurred simultaneously with the enhancements in energetic particle fluxes observed at geosynchronous altitude. Figure \ref{figS1}B shows the solar wind magnetic field. The magnetic field suggests an interplanetary coronal mass ejection \cite<ICME,>{Lepping_90} intersected Earth's magnetosphere. From 08:29 UT to 11:40 UT the magnetic field components were highly variable, indicating a turbulent sheath region formed ahead of the ICME. A sharp change in the $B_Z$ component then occurred, which may signify a magnetic field discontinuity that separated a turbulent sheath field from the magnetic field structure of the ICME \cite<see, >{Zurbuchen_06, Kilpua_17}. From 11:30 UT to 12:10 the solar wind magnetic field slowly rotated indicating a large flux rope was embedded in the solar wind. As $B_Z$ passed through zero, $B_Y$ reached a minimum of -40 nT, demonstrating the solar wind monitor was passing through the core of the flux rope where the field is mostly axial \cite{Burlaga_88}. The $B_Y$ dominated solar wind may still contribute to significant dayside reconnection \cite{Fuselier_11}. Figure \ref{figS1}C shows the solar wind dynamic pressure. From 00:00 UT to 09:10 UT, the dynamic pressure was low, near 2-5 nPa. As the solar wind monitor passes through the ICME sheath region the solar wind thermal pressure increases and becomes more turbulent. The dynamic pressure continued to increase from 09:10 UT to 15:30 UT, reaching a maximum of 30 nPa, coinciding with a sharp change in $B_Y$. Figure \ref{figS1}D shows the SMR and SYM-H indices. A common two minimum profile is seen, likely driven by the two instances of southward solar wind $B_Z$. The red curve, SMR-18, shows the lowest minima at around 11:30 UT and 18:10 UT, suggesting that the ring current was strongest around 18 MLT.

\noindent\textbf{S2: VNERX Event Validation: Simultaneous Vx and Bz Reversal Signatures}

In Figure \ref{figS2} we concentrate on the magnetic field and velocity observations from THEMIS of the three VNERX events found and establish the coherent $B_Z$ and $V_X$ behavior that implies these periods were VNERX detections. We note that the ion bulk velocity perpendicular to the magnetic field $V_{\perp,x}$ was found to be very similar to the ion bulk velocity $V_{X,GSM}$, thus we only show the latter here. 

For VNERX \#1 there are multiple $B_Z$ crossing times, however the fast tailward flows only occur between 14:25:30 UT - 14:28:00 UT and thus only the $B_Z$ crossing that occurred in that period was deemed a VNERX site passage. During VNERX \#2 multiple negative $B_Z$ excursions occurred that were all associated with a tailward fast flow signature. These were likely signatures of tailward jets from the VNERX site located earthward of THD. For VNERX \#3, between 19:20 UT and 19:25 UT THD observed several negative $B_Z$ spikes that all occurred during strong tailward flows. The $B_Z$ zero crossing associated with the VNERX site crossing occurred at 19:27, when THD was close to the southern magnetotail lobe. After the crossing THD observed fast earthward flows, thus even though THD was far from the plasma sheet we are confident a VNERX site passed tailward of THD.

\noindent\textbf{S3: VNERX Case Study}

Between 16:30 UT - 17:30 UT THA was located at [X,Y,Z] = [-12.6, 3.02, -1.11] $R_E$. Figure \ref{figS3} presents the magnetic field and plasma measurements from THA, KOMPSAT, and Arase during this period. The Supporting Information S7 and S9 presents the same analysis as Figure \ref{figS3} for VNERX \#1 and \#3 respectively and Figures S8 and S10 present the 2D XZ ion velocity distribution cuts during the observation of fast tailward reconnection ejecta for VNERX \# 1 and \# 3 respectively.

THA observed four instances of fast tailward flows, shown in Figure \ref{figS3}B, demarcated by the first four vertical black dashed lines that show the times of peak $|V_X|$. These times align with periods of total magnetic field strength decrease shown in Figure \ref{figS3}A suggesting these flows were seen when THA approached the neutral sheet. We confirm this by calculating the plasma beta in Figure \ref{figS3}D which demonstrates the fast tailward flows occur during peaks in the plasma beta that reach or exceed one. Figure \ref{figS3}C shows $E_Y$ calculated from $-\mathbf{V}\times \mathbf{B}$ and observed from the EFI instrument. Both curves show intensification of $E_Y$ that aligns with fast tailward flows, the red dashed horizontal line denotes the 5 mV/m level. These observations suggest THA was within the plasma sheet, above the neutral plane and witnessed four instances of tailward fast flows likely from an earthward reconnection site. Figure \ref{figS3}E and F shows the high- and low-energy ion energy flux measured from the SST and ESA instruments respectively. A cold ion beam indicative of ionospheric outflow is observed in the low energies. The periods of fast tailward flows however show an absence of that beam population, indicating that these fast flows are due to a hotter, plasma sheet population. After the fourth fast flow (17:07:40 UT), THA observes a reversal of ion bulk velocity within five minutes. At the reversal time, 17:08 UT, THA is not exactly at the neutral sheet, so the reversal of $B_Z$ associated with the flow reversal is small. The concomitant $V_X$ and $B_Z$ reversals alongside the previous enhancements in $E_Y$ suggests THA was sensing the outflows of a VNERX. The tailward traveling VNERX site likely passed below THA at the reversal time.

During this period KOMPSAT was situated at geosynchronous altitudes in the post-midnight sector. Figure \ref{figS3}G shows the KOMPSAT magnetic field observations. $B_X$ implies the post-midnight geomagnetic field was still stretched for over 10 minutes after THA observed the VNERX signatures. At $\sim$17:21 UT KOMPSAT observed a dipolarization. Figure \ref{figS3}H and I show the electron and proton number fluxes. At 17:20 UT KOMPSAT measures strong enhancements in the electron fluxes and slight enhancements in the proton fluxes. The observations demonstrate that geosynchronous activity indicative of ring current enhancement occurred alongside the VNERX observation.

Arase at this time was in the pre-midnight sector at [X,Y,Z] = [-3.08, 2.56, -1.85] $R_E$. Figure \ref{figS3}J shows the magnetic field observations. $B_X$ shows the field is relative constant at -400 nT, but decreases in magnitude starting at 17:08 UT while $B_Z$ increases. The observations indicate the magnetic field was dipolarizing. Figure \ref{figS3} K and L show the ion and electron number flux respectively. Both panels show an enhancement of number flux at the time of dipolarization. We infer from the temporal correlation of the THA VNERX observations and Arase flux enhancements that VNERX sites were ongoing and may have contributed to these inner-magnetospheric flux enhancements.

THD was located tailward and THE was dawnward of THA at [X,Y,Z] = [-12.70,2.36,-1.95] and [-12.58,1.30,-1.58] respectively. The observations from these spacecraft contribute to our picture of the spatial profile of fast flows in the very-near-Earth region during this period. Figure \ref{figS4} shows the magnetic field, ion bulk flow, electric field, and energy flux measured by THD and THE.

Figure \ref{figS4}A and B show the magnetic field observations of THD and THE, demonstrating both spacecraft frequently observed the plasma sheet and approached the close to the neutral sheet. Figure \ref{figS4}C and D show the bulk ion velocities measurements for THD and THE respectively. Over a 20 minute period both spacecraft see fast tailward flows near -500 km/s. The period when THD and THE see fast tailward flows overlap the period when THA observes the flow reversal. While THA was observing fast earthward flows, near $\sim$17:11 UT, THD and THE were tailward of THA and observed fast tailward flows. Assuming the flows originate from a single VNERX site, we infer that at $\sim$17:11 UT the VNERX site was situated tailward of THA but earthward of THD. THD and THE see the flow reversal 12 minutes later at $\sim$17:23 UT. Figure \ref{figS4}E and F display $E_Y$, and we find that during the period of strong tailward flows $E_Y$ was often above 20 mV/m. The agreement between measured $E_Y$ and estimated $E_Y$ from $-\mathbf{V} \times \mathbf{B}$ in Figure \ref{figS4}E demonstrates that the $E_Y$ is due to convective motion, magnetic flux moving tailward. Figure \ref{figS4}G shows plasma beta for THE and THD in red and blue respectively. During the fast flow periods both spacecraft were observing high plasma beta regimes, and thus were situated deep within the plasma sheet, close to the neutral sheet. Figure \ref{figS4}H and I present the ion energy flux for THD and THE. During the fast tailward periods both spacecraft observed a hot plasma and during periods when the flow is small we see a cold ion beam occur. The slow flow periods are times when the spacecraft observed large $B_X$ and small plasma beta, thus imply the spacecraft left the plasma sheet where cold ionospheric outflow was observed.

We expect VNERX to occur within a thin current sheet, and here we estimate how thin the current sheet was at THEMIS during the VNERX site passage. We use a 1D Harris current sheet fit \cite{Harris_62} to estimate the half-thickness of the current sheet. The fit requires simultaneous measurements of the magnetic field from two spacecraft and their respective $Z_{GSM}$ locations. Figure \ref{figS5} shows the estimate, the red, blue, and black curves gives the estimate using THD and THE, THA and THE, and THA and THD respectively. The vertical dashed line denotes the approximate time of VNERX passage and the horizontal dashed line denotes the 1 $R_E$ level. During this time, THA and THE were both observing positive $B_X$, thus were on the same side of the neutral line, and are not an adequate pair for estimating current sheet thickness. However, the remaining pairs (THA-THD and THD-THE) were on opposite sides of the neutral sheet and thus are appropriate for estimating the sheet thickness. The red and black curves show that prior to the VNERX passage, both estimates show the current sheet was thinner than 1 $R_E$. These estimates illustrate why VNERX are so elusive. Since the inter-spatial $Z_{GSM}$ separation of the THEMIS inner probes is 1 $R_E$, VNERX events embedded within a current sheet less than 1 $R_E$ can be easily missed.

This case study demonstrated strong particle enhancements at geosynchronous and sub-geosynchronous altitudes at times shortly following a VNERX observation. The implication of these results is that VNERX ejecta can result in particle injections deep in the inner magnetosphere. The observations show the magnetic field at both KOMPSAT and Arase was highly stretched, indicating the plasma sheet had advanced earthward significantly. Such an environment is conducive to reconnection, and thus it is reasonable to suggest reconnection sites are scattered throughout the very-near-Earth environment and play an important role in introducing energetic particles to the region. Two more VNERX events were observed during the storm main phase and in the next section we aim to show what the common traits of VNERX events are through multi-case study analysis.

\clearpage

\noindent\textbf{S4: Multi-case Analysis: Characterizing the Features of VNERX}

This study is concerned with understanding the frequency and geo-effectiveness of VNERX. In this section we demonstrate that the VNERX observed had the hallmarks of x-line reconnection and quantify their intensity. To that end, here we describe the characteristics of the VNERX observed. We show that the VNERX exhibit the classical quadrupolar Hall $B_Y$ and $E_Y$ signatures. The reconnection electric field $E_Y$ of 5 - 40 mV/m was seen throughout all events. Strong inflows greater than 200 km/s were frequently seen during VNERX passage.

Figure \ref{figS6} shows the electromagnetic fields associated with the VNERX sites. Figure \ref{figS6}A shows the y-component of the electric field observed by THA and THD over the same period as above as a function of $B_Z$ and $V_{i,X}$. We filtered the data to only include times when the magnitude of the ion velocity was greater than $|200|$ km/s to restrict ourselves to times of reconnection ejecta observation. We see that the data primarily populates the top right and bottom left quadrants and $E_Y$ is generally positive in these regions, demonstrating the expected reconnection electric field signature in the x-line reconnection geometry.

Figure \ref{figS6}B shows the y-component of the magnetic field observed by THA and THD over the ten hour period 12:00 UT - 22:00 UT as a function of $B_X$ and $B_Z$. We've filtered the data points to only include magnetic field data observed during times when the y-component of the electric field was greater than 5 mV/m to restrict ourselves to only VNERX passage times. Since THE did not have EFI data during this time, magnetic field from this spacecraft was not included. The data shows a positive $B_Y$ in the first and third quadrants and negative $B_Y$ in the second and fourth quadrants. We interpret this as the detection of the quadrupolar $B_Y$ associated with an X-line reconnection geometry \cite{Sonnerup_79}.

The electromagnetic fields described here match our expectations of the field directions for x-line reconnection \cite{Hesse_Cassak_2020}. Comparing to the average electromagnetic field strengths seen in ion diffusion regions of distant x-lines described in \citeA{Eastwood_2010}, we find that: 1) the quadrupolar $B_Y$ fields are larger than the average values seen at more distant x-lines. 2) The reconnection electric field $E_Y$ is generally stronger than the typically less than 10 mV/m fields seen in the ion diffusion region of distant tail x-lines. The strength of the Hall and reconnection fields found in this study are consistent with those demonstrated by \citeA{Beyene_Angelopoulos_24}. These differences are a result of the stronger magnetic field available in the very-near-Earth environment. These results also demonstrate that the amount of flux throughput is stronger for VNERX than reconnection downtail. This becomes an important factor for ring current energization because the flux tubes generated by VNERX are more likely to access the ring current region since they are no longer blocked by the near-Earth transition region that frequently impedes flux tubes generated from more distant reconnection \cite{Sergeev_12}. Consequently, energy throughput may be higher for VNERX than near-Earth reconnection, implying VNERX may play a strong role in carrying energy into the inner-magnetosphere and ring current. 

\clearpage


%
%


%
%
\bibliography{VNERX_references}
%
%
%


%
%
%
%
%

%
%
\end{article}
\clearpage


%
%

\begin{figure}
\noindent\includegraphics[width=0.99\textwidth]{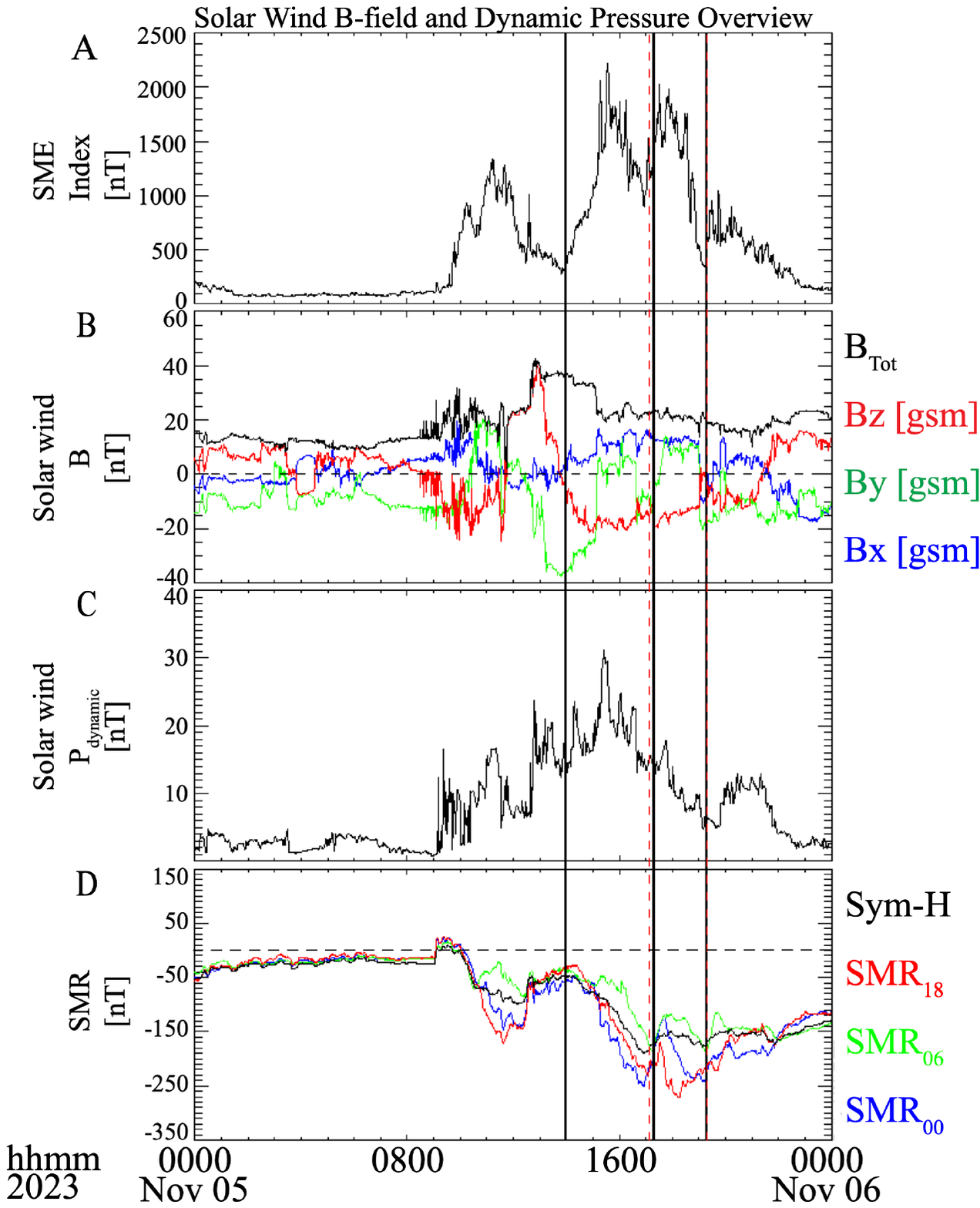}
\end{figure}

\begin{figure}
\caption{ Overview of solar wind activity and magnetic indices during November 05 2023. The black vertical lines denote when KOMPSAT observed dispersionless injections in the electron fluxes, and the red dashed vertical lines denote times when Arase observed simultaneous magnetic field dipolarizations and proton number flux enhancements. A) The SME index.  B) The solar wind magnetic field measured at the L1 Lagrange point and time-lagged to the nose of Earth's bowshock. C) Solar wind dynamic pressure. D) The SMR and SYM-H indices. The red, green, and blue curves are the deviations of the horizontal magnetic field in the MLT sector wedges centered on MLT 18 hr, 06 hr, and 00 hr, respectively. The black curve gives the average deviation across all MLTs. The horizontal black dashed line denotes 0 nT.}
\label{figS1}
\end{figure}


\begin{figure}
\noindent\includegraphics[width=\textwidth]{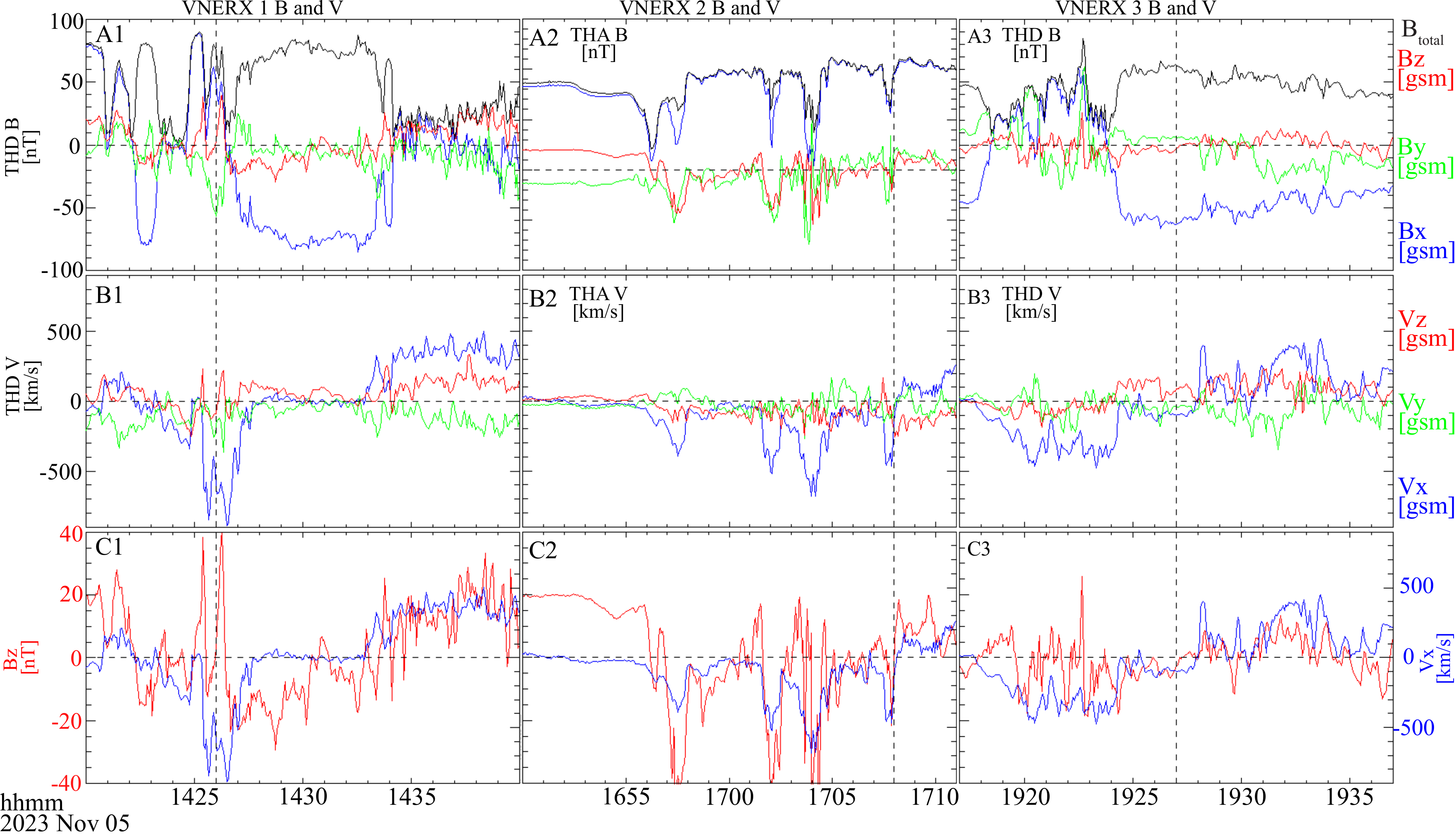}
\caption{The observed magnetic field and velocity profiles from THEMIS of the three VNERX events. Top panels (a1 - a3) show the magnetic field in GSM coordinates and the magnitude in black. The middle panels (b1 - b3) show the ion bulk velocity. The bottom panels (c1 - c3) show $B_Z$ in red and $V_X$ in blue. Note the left y-axis for the bottom panels corresponds to the $B_X$ data, and the right y-axis corresponds to the $V_X$ data. The verticle dashed lines denote the $B_Z$ zero crossing time associated with the VNERX site.}
\label{figS2}
\end{figure}

\begin{figure}
\noindent\includegraphics[width=\textwidth]{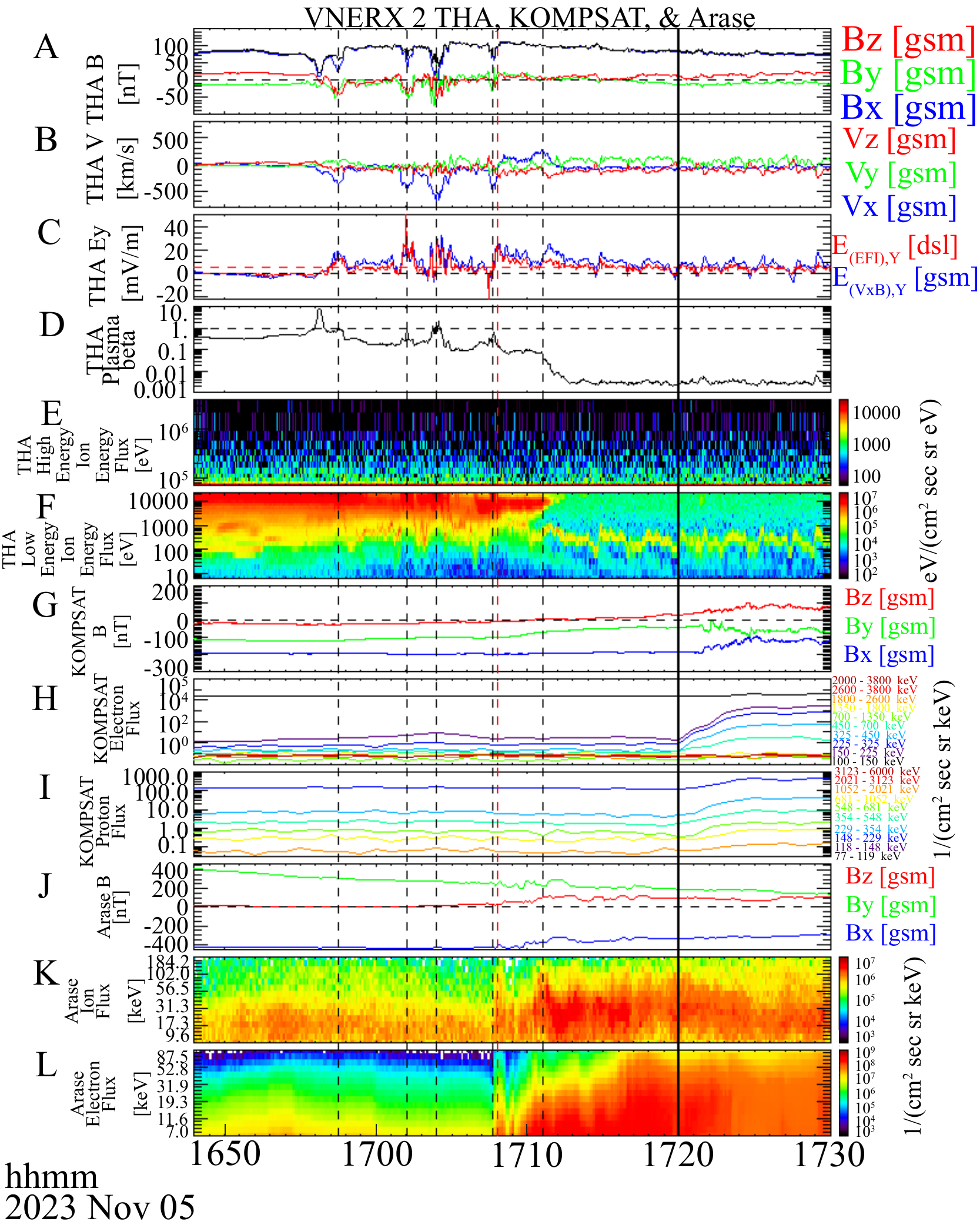}
\end{figure}

\begin{figure}
\caption{VNERX 2 THA, KOMPSAT, and Arase observations. The solid black vertical line denotes the time KOMPSAT observed the dispersionless injection. The red dashed vertical line denotes the time the Arase MEP-I observed an increase in the ion energy flux. The dashed vertical black lines denote times THA saw fast flows. All vector quantities are in GSM coordinates unless otherwise specified. A) Magnetic field observed by THA. B) Ion bulk velocity moment derived from ESA ion distribution. C) Y-component of electric field observed at THA in GSM and DSL coordinates. D) Plasma beta $\beta = \frac{P_{thermal}}{P_{magnetic}}$. E) High energy ion energy fluxes measured by THA SST. F) Low energy ion energy fluxes measured by THA ESA. G) Magnetic field observed by KOMPSAT. H) Electron number fluxes observed by KOMPSAT. I) Proton number fluxes observed by KOMPSAT. J) Magnetic field observed by Arase. K) Proton number flux observed by Arase MEP-I. L) Electron number flux observed by Arase MEP-E.}
\label{figS3}
\end{figure}

\begin{figure}
\noindent\includegraphics[width=\textwidth]{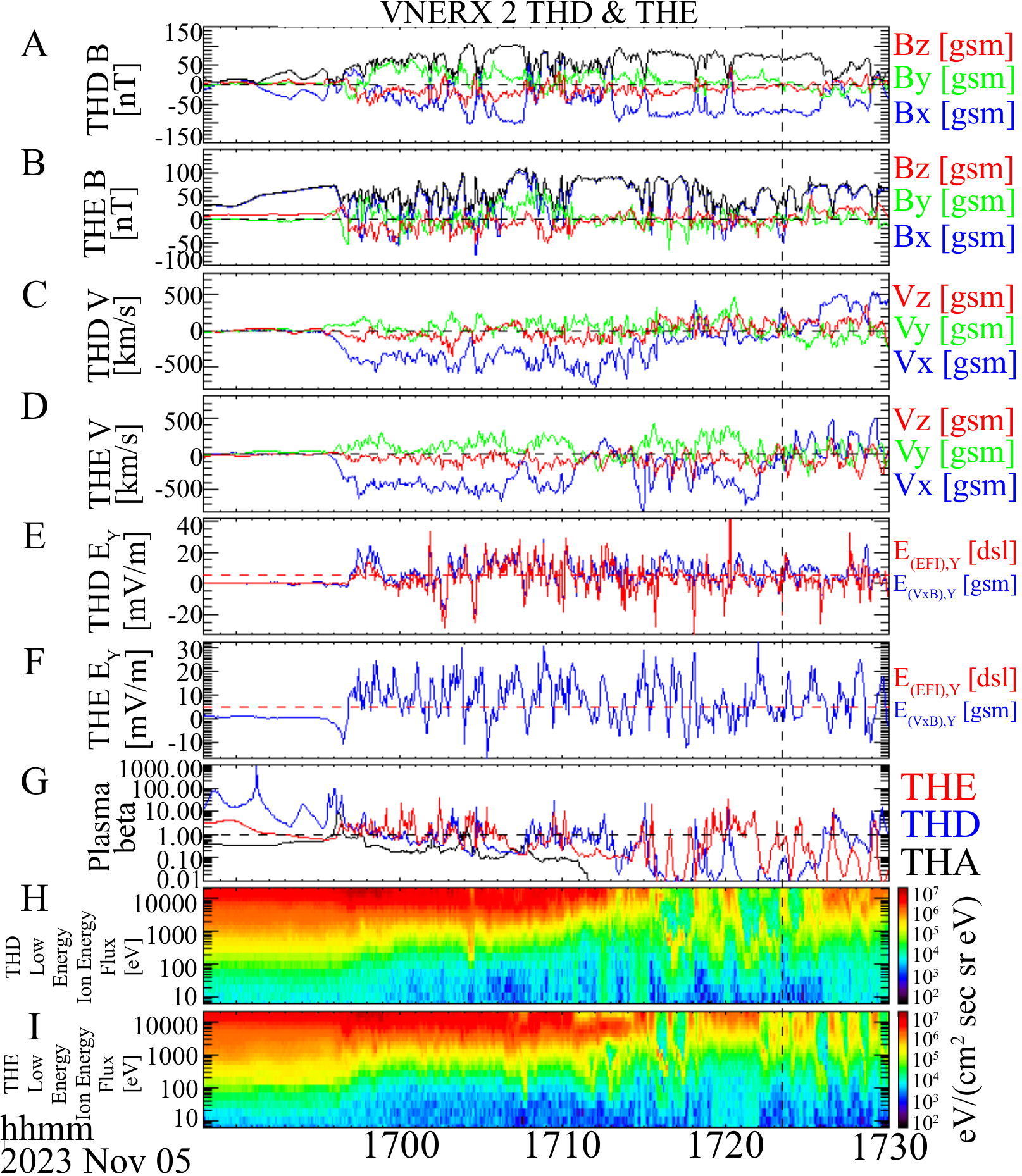}
\end{figure}

\begin{figure}
\caption{Overview of VNERX 2 observations from THD and THE. The vertical dashed line at 17:23 UT denotes the period when THD and THE observed $V_X$ reverse from tailward to earthward. A) THD magnetic field. B) THE magnetic field. C) THD ion bulk velocity. D) THE ion bulk velocity. E) THD $E_Y$ from the EFI instrument and estimated from $-\mathbf{V}×\mathbf{B}$. F) THE $E_Y$ estimated from $-\mathbf{V} × \mathbf{B}$ G) plasma beta for THA, THD, and THE. H) THD low-energy ion energy flux. I) THE low energy ion energy flux.} 
\label{figS4}
\end{figure}

\begin{figure}
\noindent\includegraphics[width=\textwidth]{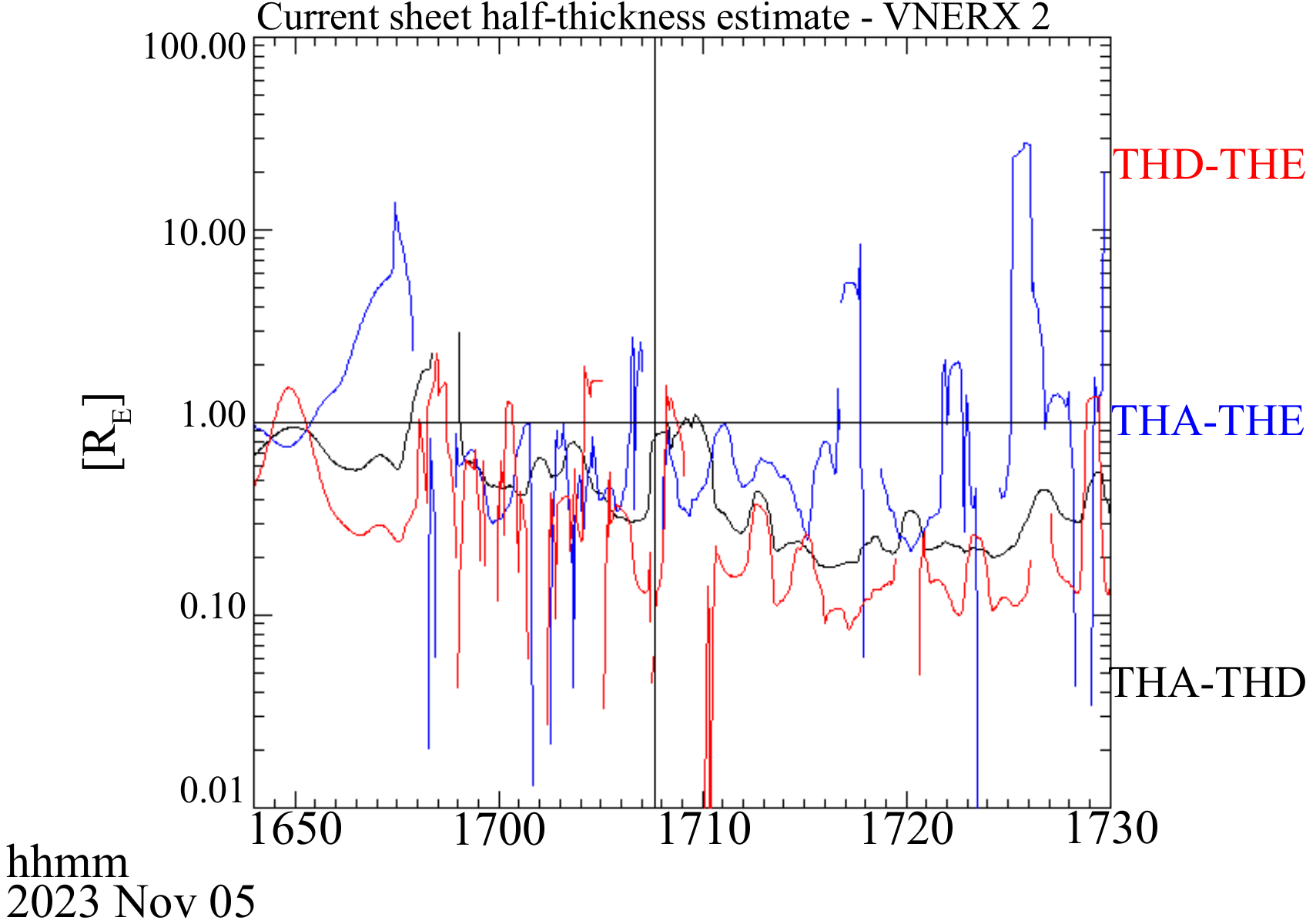}
\caption{Harris current sheet thickness estimate for VNERX \#2. The different colored curved present the estimate using different pairs of spacecraft. The red curve used THD and THE, the blue curve used THA and THE, and the black curve used THA and THD. }
\label{figS5}
\end{figure}

\begin{figure}
\noindent\includegraphics[width=.96\textwidth]{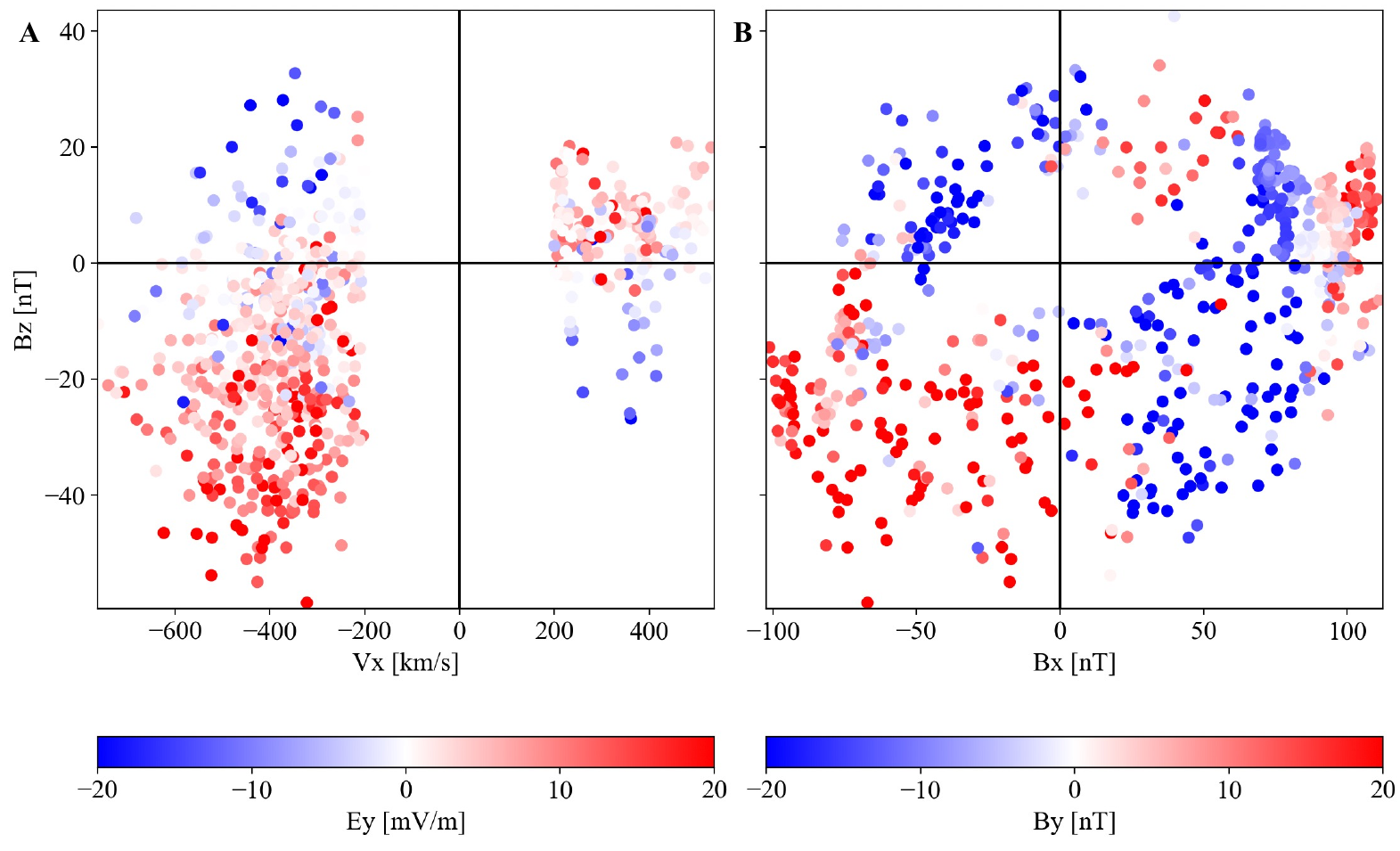}
\caption{Electromagnetic fields observed with VNERX passage. A) Reconnection electric field $E_Y$ in DSL coordinates as a function of $V_X$ and $B_Z$. B) Hall magnetic field $B_Y$ as a function of $B_X$ and $B_Z$.}
\label{figS6}
\end{figure}

\begin{figure}
\centering
\includegraphics[width=0.9\textwidth]{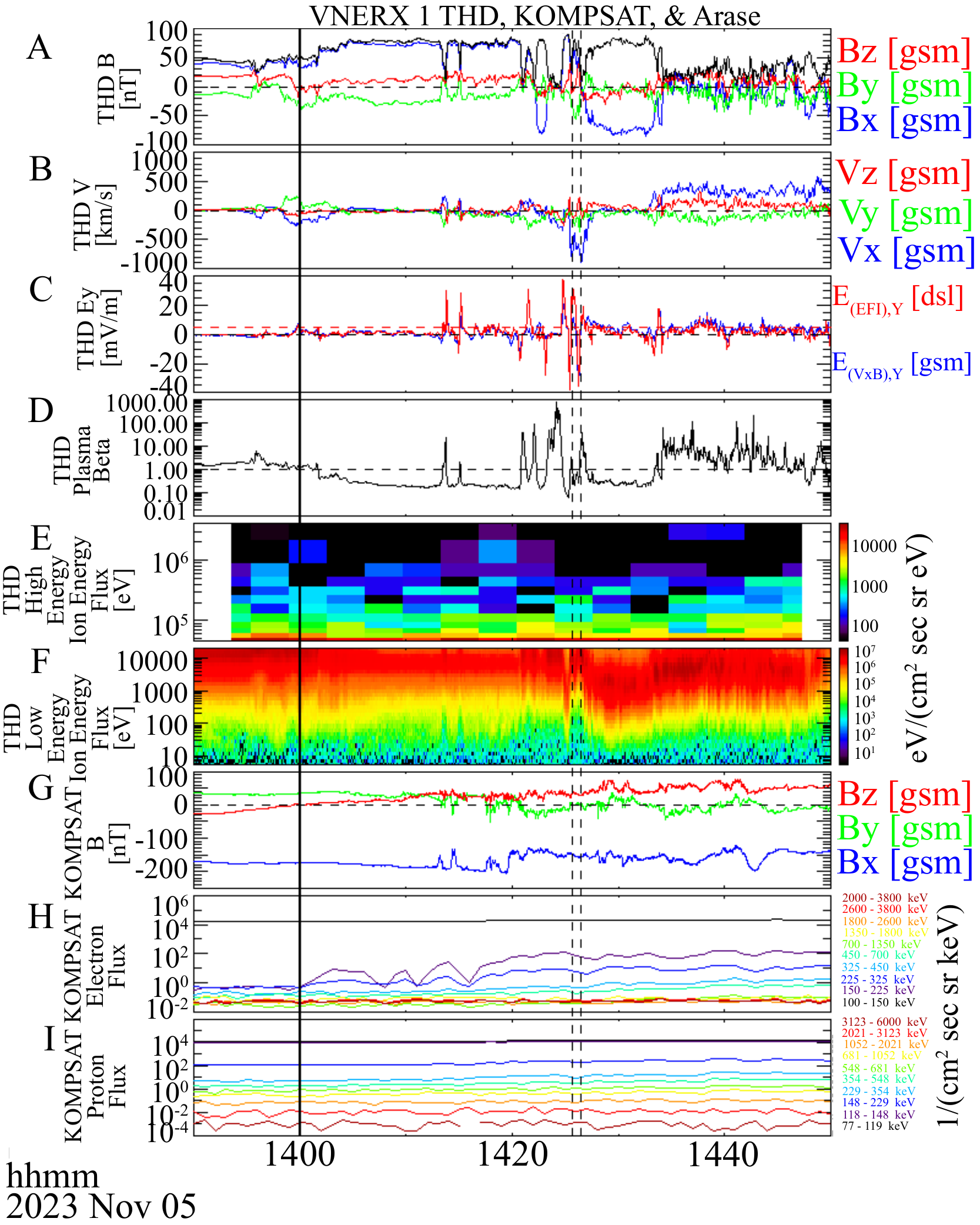}
\end{figure}

\begin{figure}
\caption{VNERX 1 THD, and KOMPSAT observations. The solid black vertical line denotes the time KOMPSAT observed the dispersionless injection. The dashed vertical black lines denote times THD saw fast flows. All vector quantities are in GSM coordinates unless otherwise specified. A) Magnetic field observed by THD. B) Ion bulk velocity moment derived from ESA ion distribution. C) Y-component of electric field observed at THD in GSM and DSL coordinates. D) Plasma beta $\beta = \frac{P_{thermal}}{P_{magnetic}}$. E) High energy ion energy fluxes measured by THD SST. F) Low energy ion energy fluxes measured by THD ESA. G) Magnetic field observed by KOMPSAT. H) Electron energy fluxes observed by KOMPSAT. I) Proton energy fluxes observed by KOMPSAT.}
\label{figS7}
\end{figure}

\begin{figure}
\centering
\includegraphics[width=0.9\textwidth]{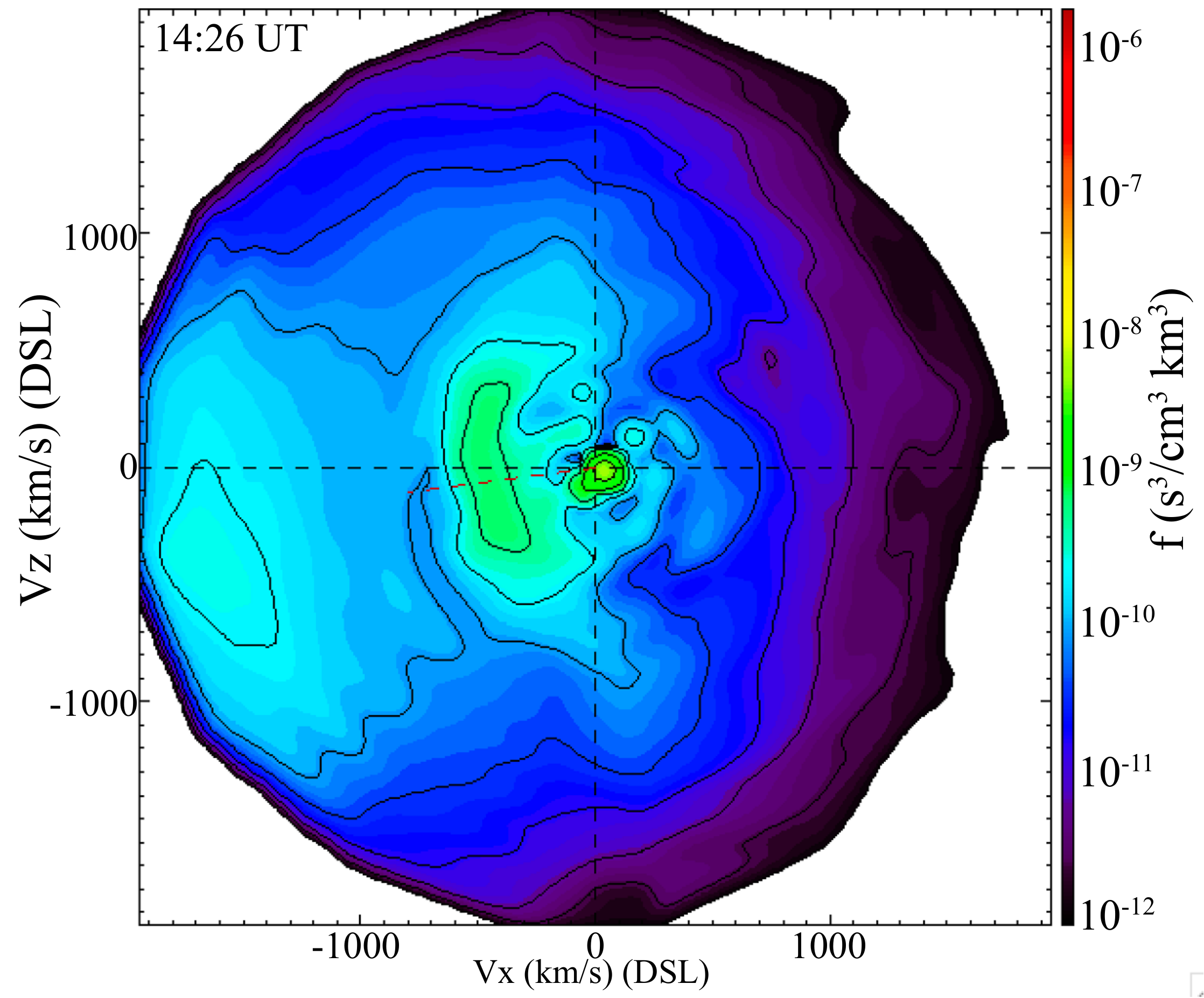}
\end{figure}

\begin{figure}
\caption{2D XZ velocity cut in DSL coordinates for THD during a fast tailward flow for VNERX 1. The vertical and horizontal dashed lines denote zero for $V_X$ and $V_Z$ respectively.}
\label{figS8}
\end{figure}

\begin{figure}
\centering
\includegraphics[width=0.9\textwidth]{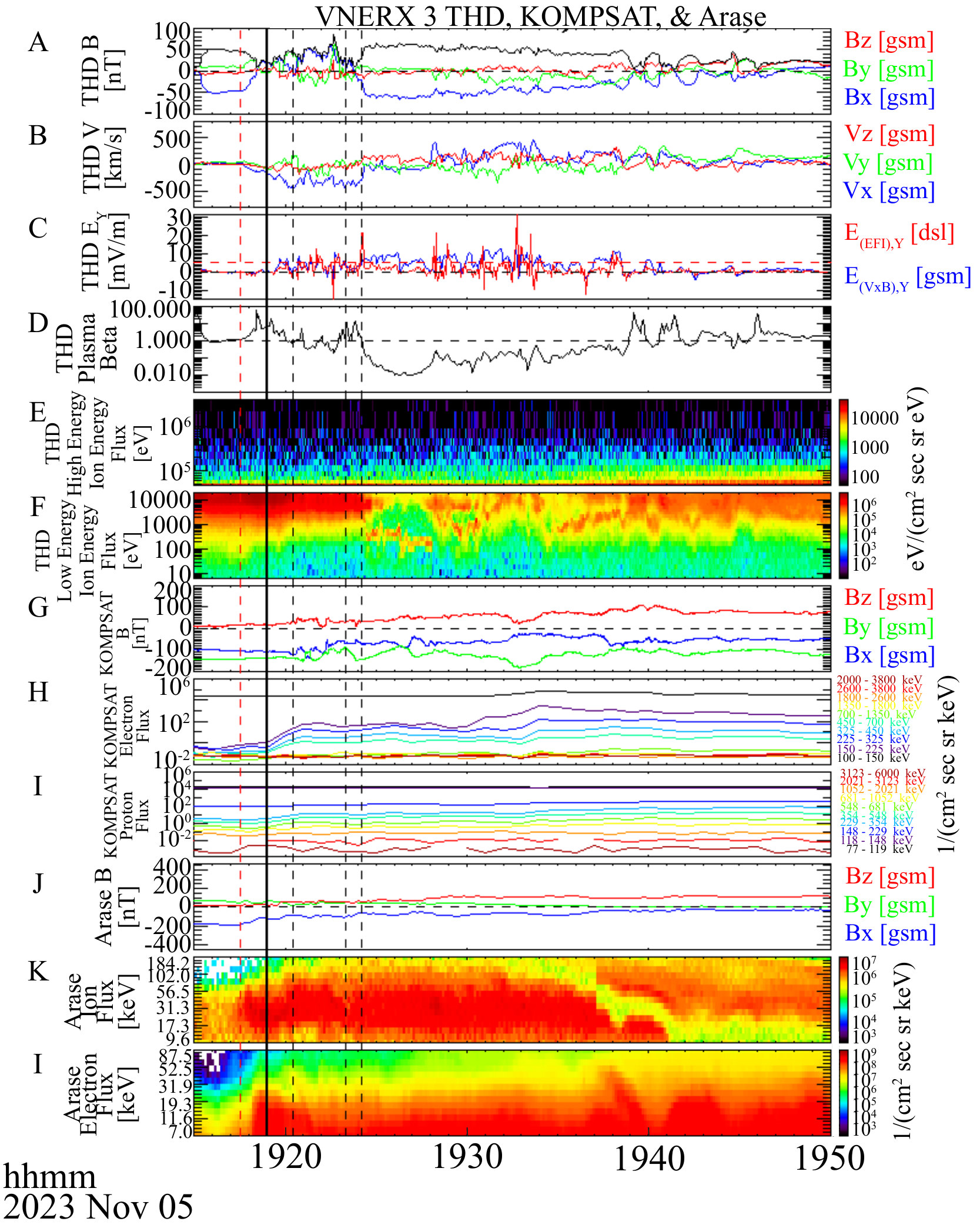}
\end{figure}

\begin{figure}
\caption{VNERX 3 THD, KOMPSAT, and Arase observations. The solid black vertical line denotes the time KOMPSAT observed the dispersionless injection. The red dashed vertical line denotes the time the Arase MEP-I observed an increase in the ion energy flux. The dashed vertical black lines denote times THD saw fast flows. All vector quantiies are in GSM coordinates unless otherwise specified. A) Magnetic field observed by THD. B) Ion bulk velocity moment derived from ESA ion distribution. C) Y-component of electric field observed at THD in GSM and DSL coordinates. D) Plasma beta $\beta = \frac{P_{thermal}}{P_{magnetic}}$. E) High energy ion energy fluxes measured by THD SST. F) Low energy ion energy fluxes measured by THD ESA. G) Magnetic field observed by KOMPSAT. H) Electron energy fluxes observed by KOMPSAT. I) Proton energy fluxes observed by KOMPSAT. J) Magnetic field observed by Arase. K) Proton energy flux observed by Arase MEP-I. L) Electron energy flux observed by Arase MEP-E.}
\label{figS9}
\end{figure}

\begin{figure}
\centering
\includegraphics[width=0.9\textwidth]{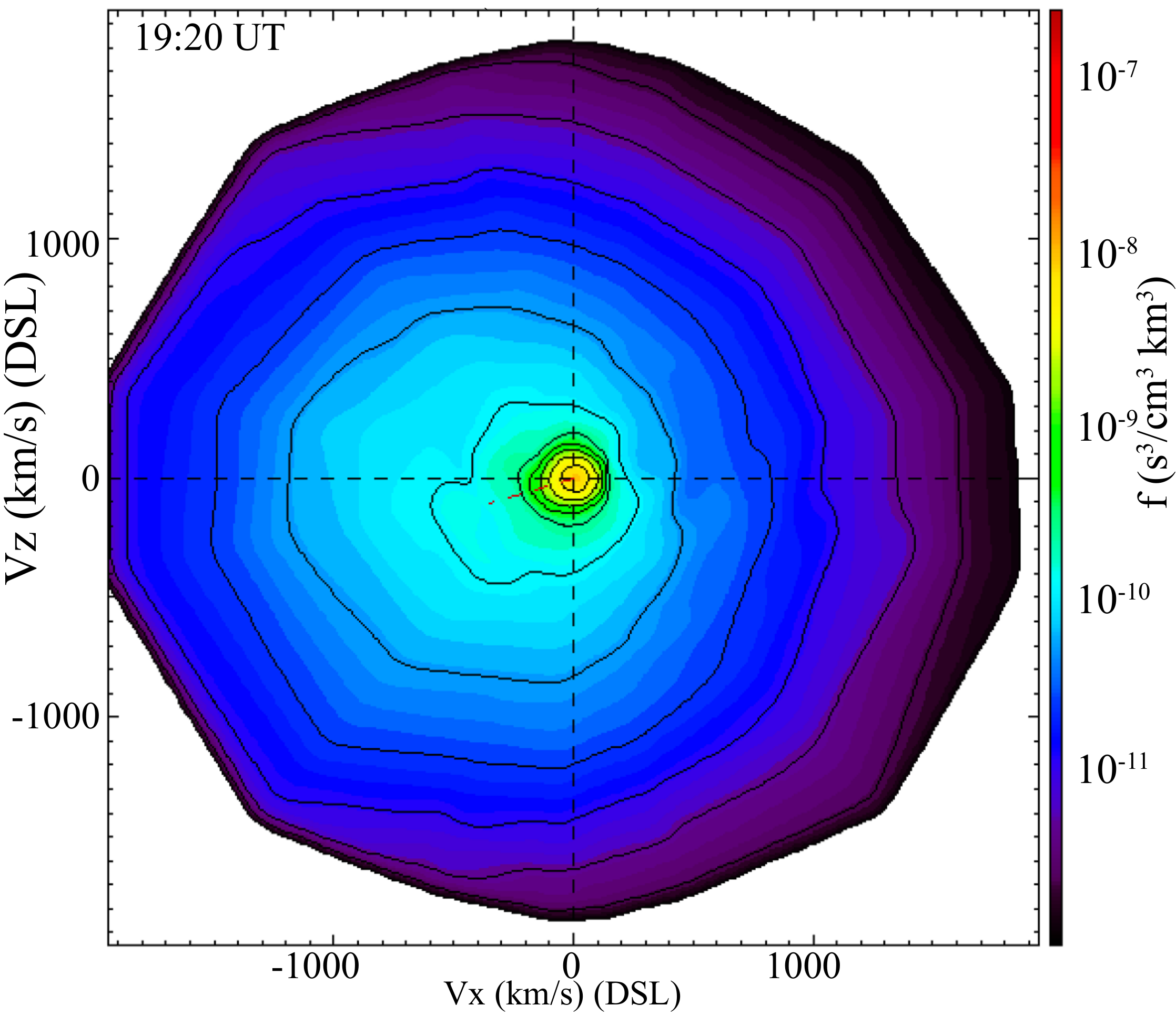}
\end{figure}

\begin{figure}
\caption{2D XZ velocity cut in DSL coordinates for THD during a fast tailward flow for VNERX 3. The vertical and horizontal dashed lines denote zero for $V_X$ and $V_Z$ respectively.}
\label{figS10}
\end{figure}